\documentclass[fleqn,usenatbib]{mnras}

\usepackage{newtxtext,newtxmath}

\usepackage[T1]{fontenc}
\usepackage{ae,aecompl}

\usepackage{myaasmacros}
\usepackage{graphicx}	
\usepackage{booktabs,tabularx,graphicx,amsmath,wasysym,makecell,hhline,enumitem,fancyvrb,gensymb,multirow,ulem,pifont,adjustbox,dblfloatfix,float}
\usepackage[dvipsnames]{xcolor}
\usepackage[printonlyused,nohyperlinks,nolist]{acronym} 
\usepackage{pdflscape, afterpage}

\definecolor{Halo1}{HTML}{305FEA}
\definecolor{Halo2}{HTML}{00A982}
\definecolor{Halo3}{HTML}{7F0579}
\definecolor{Halo4}{HTML}{00D1D4}
\definecolor{Halo5}{HTML}{EF1414}
\definecolor{GM}{HTML}{EAA209}
\definecolor{GM2}{HTML}{000000}

\title[The shape of dark matter haloes in the faintest galaxies]{EDGE: The shape of dark matter haloes in the faintest galaxies}

\author{M. D. A. Orkney; ...]{Matthew D. A. Orkney$^1$, Ethan Taylor, Justin I. Read$^{1}$\thanks{E-mail: justin.inglis.read@gmail.com}, Martin P. Rey, A. Pontzen, Oscar Agertz, Stacy Y. Kim and Maxime Delorme}\\
$^1${\small Department of Physics, University of Surrey, Guildford, GU2 7XH, Surrey, UK}\\
}

\author[Matthew D. A. Orkney et al.] 
{Matthew D. A. Orkney,$^{1,2,3}$\thanks{E-mail: morkney@icc.ub.edu} Ethan Taylor,$^{1}$ Justin I. Read,$^{1}$ Martin P. Rey,$^{4}$ Andrew Pontzen,$^{5}$ \newauthor Oscar Agertz,$^{6}$ Stacy Y. Kim,$^{1}$ and Maxime Delorme$^{7}$
\\
$^1$Department of Physics, University of Surrey, Guildford, GU2 7XH, United Kingdom\\
$^{2}$Institut de Ci\`{e}ncies del Cosmos (ICCUB), Universitat de Barcelona, Mart\'{i} i Franqu\`{e}s 1, E-08028 Barcelona, Spain\\
$^{3}$Institut d'Estudis Espacials de Catalunya (IEEC), E-08034 Barcelona, Spain\\
$^4$Sub-department of Astrophysics, University of Oxford, Keble Road, Oxford OX1 3RH, United Kingdom\\
$^5$Department of Physics and Astronomy, University College London, London WC1E 6BT, UK\\
$^6$Lund Observatory, Division of Astrophysics, Department of Physics, Lund University, Box 43, SE-221 00 Lund, Sweden\\
$^7$Département d'Astrophysique/AIM, CEA/IRFU, CNRS/INSU, Université Paris-Saclay, 91191 Gif-Sur-Yvette, France
}
\date{Accepted 2023 August 17. Received 2023 July 23; in original form 2023 February 24}

\pubyear{2023}

\begin{document}
\label{firstpage}
\pagerange{\pageref{firstpage}--\pageref{lastpage}}
\maketitle

\acrodef{AGB}{Asymptotic Giant Branch}
\acrodef{AHF}{Amiga Halo Finder}
\acrodef{AGN}{Active Galactic Nuclei}
\acrodef{AMR}{Adaptive Mesh Refinement}
\acrodef{BBN}{Big Bang Nucleosynthesis}
\acrodef{CC}{cusp-core problem}
\acrodef{CDM}{Cold Dark Matter}
\acrodef{FDM}{Fuzzy Dark Matter}
\acrodef{CMB}{Cosmic Microwave Background}
\acrodef{dE}{Dwarf Elliptical galaxy}
\acrodef{dIrr}{Dwarf Irregular galaxy}
\acrodef{DM}{Dark Matter}
\acrodef{DMO}{Dark Matter Only}
\acrodef{dSph}{Dwarf Spheroidal galaxy}
\acrodef{EDGE}{Engineering Dwarfs at Galaxy formation's Edge}
\acrodef{FTT}{Fully Threaded Tree}
\acrodef{GC}{Globular Cluster}
\acrodef{GM}{Genetic Modification}
\acrodef{HPC}{High Performance Computing}
\acrodef{ICs}{Initial Conditions}
\acrodef{IGM}{Inter-Galactic Medium}
\acrodef{IMF}{Initial Mass Function}
\acrodef{ISM}{Inter-Stellar Medium}
\acrodef{JWST}{James Webb Space Telescope}
\acrodef{LCDM}[$\Lambda$CDM]{Lambda Cold Dark Matter}
\acrodef{LG}{Local Group}
\acrodef{LHC}{Large Hadron Collider}
\acrodef{Lya forest}{Lyman-Alpha forest}
\acrodef{MACHO}{MAssive Compact Halo Object}
\acrodef{MCMC}{Markov Chain Monte Carlo}
\acrodef{MoND}{Modified Newtonian Dynamics}
\acrodef{MS}{missing satellite problem}
\acrodef{MW}{Milky Way}
\acrodef{PM}{Particle Mesh}
\acrodef{PoS}{planes of satellites problem}
\acrodef{PP}{Particle Particle}
\acrodef{SC}{Star Cluster}
\acrodef{SKA}{Square Kilometer Array}
\acrodef{SMHM}{Stellar Mass Halo Mass}
\acrodef{SNe}{Supernovae}
\acrodef{TBTF}{Too Big To Fail problem}
\acrodef{WDM}{Warm Dark Matter}
\acrodef{WIMP}{Weakly Interacting Massive Particle}
\acrodef{SIDM}{Self Interacting Dark Matter}

\begin{abstract}
Collisionless \ac{DMO} structure formation simulations predict that \ac{DM} haloes are prolate in their centres and triaxial towards their outskirts. The addition of gas condensation transforms the central \ac{DM} shape to be rounder and more oblate. It is not clear, however, whether such shape transformations occur in `ultra-faint' dwarfs, which have extremely low baryon fractions.
We present the first study of the shape and velocity anisotropy of ultra-faint dwarf galaxies that have gas mass fractions of $f_{\rm gas}(r<R_{\rm half}) < 0.06$. These dwarfs are drawn from the \ac{EDGE} project, using high resolution simulations that allow us to resolve \ac{DM} halo shapes within the half light radius ($\sim 100$\,pc).
We show that gas-poor ultra-faints ($M_{\rm 200c} \leqslant 1.5\times10^9$\,M$_\odot$; $f_{\rm gas} < 10^{-5}$) retain their pristine prolate \ac{DM} halo shape even when gas, star formation and feedback are included. This could provide a new and robust test of \ac{DM} models. By contrast, gas-rich ultra-faints ($M_{\rm 200c} > 3\times10^9$\,M$_\odot$; $f_{\rm gas} > 10^{-4}$) become rounder and more oblate within $\sim 10$ half light radii.
Finally, we find that most of our simulated dwarfs have significant radial velocity anisotropy that rises to $\tilde{\beta} > 0.5$ at $R \gtrsim 3 R_{\rm half}$. The one exception is a dwarf that forms a rotating gas/stellar disc because of a planar, major merger. Such strong anisotropy should be taken into account when building mass models of gas-poor ultra-faints.
\end{abstract}




\section{Introduction} \label{intro}

In the \ac{LCDM} paradigm, the mass of the Universe is predominantly made up of an invisible \ac{DM}. This \ac{DM} undergoes hierarchical gravitational collapse to form a cosmic web of structure comprising of sheets, filaments and nodes \citep{1974ApJ...187..425P, 1978MNRAS.183..341W, 1982ApJ...263L...1P}. It is in these nodes, or \ac{DM} haloes, that galaxies are born and evolve. \ac{DM} haloes are often assumed to be spherically symmetric for practical convenience \citep[e.g.][]{1996ApJ...462..563N, 2015MNRAS.451.1366P}. However, they are more accurately described by a sequence of nested, and potentially twisted, triaxial ellipsoids \citep{1986ApJ...304...15B, 1988ApJ...327..507F, 1991ApJ...378..496D, 2002ApJ...574..538J, 2006MNRAS.367.1781A}. \par

Cosmological \ac{DM} structure formation simulations show that \ac{DM} haloes tends towards a prolate (cigar) shape in their innermost radii, with the shape at higher radii becoming increasingly dependent on dynamical interactions and accretion from filaments \citep{2006PhRvD..74l3522G, 2007MNRAS.376..215B, 2008MNRAS.391.1940M, 2015MNRAS.450.2327Z, 2016MNRAS.458.4477T, 2018AAS...23135606W}. In a cosmological setting, the sum of these interactions yields a more triaxial or oblate (pancake) shape in their outskirts \citep{1988ApJ...327..507F, 2002ApJ...574..538J, 2011MNRAS.416.1377V, 2023arXiv230208853C}. This has been suggested to be a consequence of the central halo having been constructed from highly anisotropic accretion from thin filaments in the early Universe, contrasting with a more isotropic build-up of the outer halo at later times \citep{1997MNRAS.290..411T, 2011MNRAS.416.1377V, 2018MNRAS.476.1796S, 2018MNRAS.481..414G, 2023arXiv230208853C}. This interpretation is supported by \citet{2004MNRAS.354..522M} and \citet{2010MNRAS.406..744C}, which suggest that halo shape and orientation depends on the impact parameter of merging material, with radial mergers favouring prolate shapes that point in the direction of the infall and more tangential mergers favouring oblate shapes. \par

The halo shape can also be affected by the chosen cosmology. For example, the increased inter-particle interactions in a \ac{SIDM} cosmology result in a dynamical heating of their orbits. This leads to much rounder halo shapes at the halo centre, where the interactions are most frequent \citep{2001ApJ...547..574D, 2013MNRAS.430..105P, 2018MNRAS.474..746B}. However, the disparity between \ac{CDM} and \ac{SIDM} is lessened once baryonic physics is included (\citealp{2022arXiv220412502D}., and see also \citealp{2022MNRAS.516.2389V}). \ac{FDM} can also lead to a rounder \ac{DM} configuration \citep{2023ApJ...949...68D}. \par

It is well established that the halo shape is impacted by `baryonic physics' -- the addition of dissipative gas, star formation and feedback to the numerical models. Simulations that include `sub-grid' models for these effects find that an initially prolate central halo becomes rounder and more oblate and -- where a stellar and/or gas disc is present -- aligned with the disc \citep{1991ApJ...377..365K, 1994ApJ...431..617D, 2006PhRvD..74l3522G, 2008ApJ...681.1076D, 2010PhRvD..82b3531P, 2013MNRAS.429.3316B, 2016MNRAS.458.4477T, 2021MNRAS.501.5679C, 2022MNRAS.515.2681C}. The central minor-to-major axis ratio, which is often used as a measure of halo sphericity, generally increase from approximately 0.4-0.5 to 0.6-0.7. This shape-change effect is caused by a modification of the \ac{DM} particle orbits. The inner regions of a primordial prolate halo are largely comprised of \ac{DM} particles on box orbits, which can pass very close to the halo centre \citep{1985MNRAS.216..467G, 1994A&A...281..314U, 1999AJ....118.1177M}. As gas clouds cool and condense into the halo centre, their gravitational influence deflects these particles such that they become dominated by short-axis tube orbits \citep{1998MNRAS.297..177T, 1999AJ....118.1177M, 2010ApJ...720L..62K, 2017MNRAS.466.3876Z}, which favour a more oblate or rounder overall halo shape. The transformation is most efficient within $\sim20$ half light radii, where the majority of the galactic gas and stars reside \citep{2004IAUS..220..421S, 2008ApJ...681.1076D}. For similar reasons, the shape is also highly dependent on the mass-scale and redshift \citep[see][]{2008MNRAS.388.1321P, 2019MNRAS.484.5170Z}, with halo centres tending towards more oblate and rounder shapes as they become increasingly dominated by their baryons \citep[e.g.][]{2015MNRAS.453..408C, 2016MNRAS.458.4477T}. \par

Analysis of the halo shape in large volume simulations reveals a significant variation in the axial ratios across populations in multiple mass regimes \citep{2002ApJ...574..538J, 2005ApJ...629..781K, 2006MNRAS.367.1781A, 2007MNRAS.376..215B, 2019MNRAS.484..476C}. The variation in the minor-to-major axis ratios can be as great as $1\sigma\simeq0.2$. This underscores that whilst different physics schemes and mass-scales have important systematic effects on the halo shape, it is also dependent on the unique evolution and circumstances of every halo. \par

Observations suggest a diversity of halo shapes \citep[i.e.][]{2017MNRAS.464...65P}, but such measurements are challenging to make. The projected mass profiles and shapes of massive galaxies and galaxy clusters can be examined directly with gravitational lensing \citep{2010ARA&A..48...87T, 2012A&A...545A..71V, 2016MNRAS.456..870B}, but this is limited to measuring the 2D projected shape. The mean 3D shape of different object classes can only be inferred statistically, using many such measurements of the 2D shape. Tracers of the gravitational potential, such as stellar streams \citep{2001ApJ...551..294I, 2004ApJ...610L..97H, 2012MNRAS.424L..16L, 2013MNRAS.428..912D, 2013ApJ...773L...4V} and globular clusters \citep{2012MNRAS.424L..16L, 2019A&A...621A..56P}, can reveal the 3D halo shape over the radii that those tracers explore. But data of sufficient quality is only available, at present, for the Milky Way \citep[e.g.][]{2014JPhG...41f3101R}. Alternatively, edge-on galaxies provide a unique means of measuring the halo shape by analysis of the rotation and flaring in their HI discs \citep{1995AIPC..336..121O, 2008MNRAS.383.1343W, 2010A&A...515A..63O, 2017MNRAS.464...65P, 2021MNRAS.500..410L}. This can also reveal the 3D halo shape, but only in the vicinity of the gas disc. Whilst the HI flaring technique is effective in the dwarf regime, it remains difficult to detect triaxiality. \par

To summarise, the \ac{DM} halo shape contains clues about the \ac{DM} cosmology, the assembly history, and various formation mechanisms including the role of baryons. Therefore, a proper understanding of halo shapes is critical for understanding galaxy formation as a whole. Whilst the central shape of more massive galaxies becomes tightly coupled to their baryons, it is unclear how efficiently the traditional shape transformation mechanisms will function within the faintest dwarf galaxies, where gas and stellar mass fractions are proportionally far lower.  If there is a mass limit, beneath which ultra faints are unaffected by their baryonic component, then this could constrain physical models (such as the interaction cross section in \ac{SIDM} cosmologies) and inform the assumptions used in mass-modelling methods. Therefore, the \ac{DM} halo shapes of ultra faint dwarf galaxies may offer a unique perspective on galaxy formation. \par

In this paper, we investigate \ac{DM} halo shapes in the smallest galaxies in the Universe -- `ultra-faint' dwarfs. To our knowledge, this is the first dedicated study of the halo shape in galaxies of this scale. We use the \ac{EDGE} simulation suite (\citealp{rey2019, agertz2020, pontzen2020, rey2020, orkney}), with the aim of studying whether the halo shapes in gas-deficient reionisation fossils are altered by the presence of baryons. We also study the velocity anisotropy of the inner stellar and \ac{DM} particle orbits to see if these are affected by baryonic effects and to inform priors on mass modelling nearby dwarf galaxies. Finally, we compare between different assembly histories and feedback models. The paper is organised as follows:
In \S\ref{method}, we describe our simulation suite and method for calculating the halo shape. In \S\ref{results}, we present our results for the shape profile of ultra-faint dwarf galaxies, and the temporal evolution of the central shape. We present the stellar and \ac{DM} velocity anisotropy profiles in \S\ref{anisotropy}. Then, we investigate the impact of assembly history in \S\ref{GM} and feedback model in \S\ref{fblim}. We discuss the implications of our results in \S\ref{discussion}, and present our conclusions in \S\ref{conclusion}. \par

\begin{figure*}
\centering
\includegraphics[keepaspectratio, trim={0.3cm 0cm 0.2cm 0cm}, width=\linewidth]{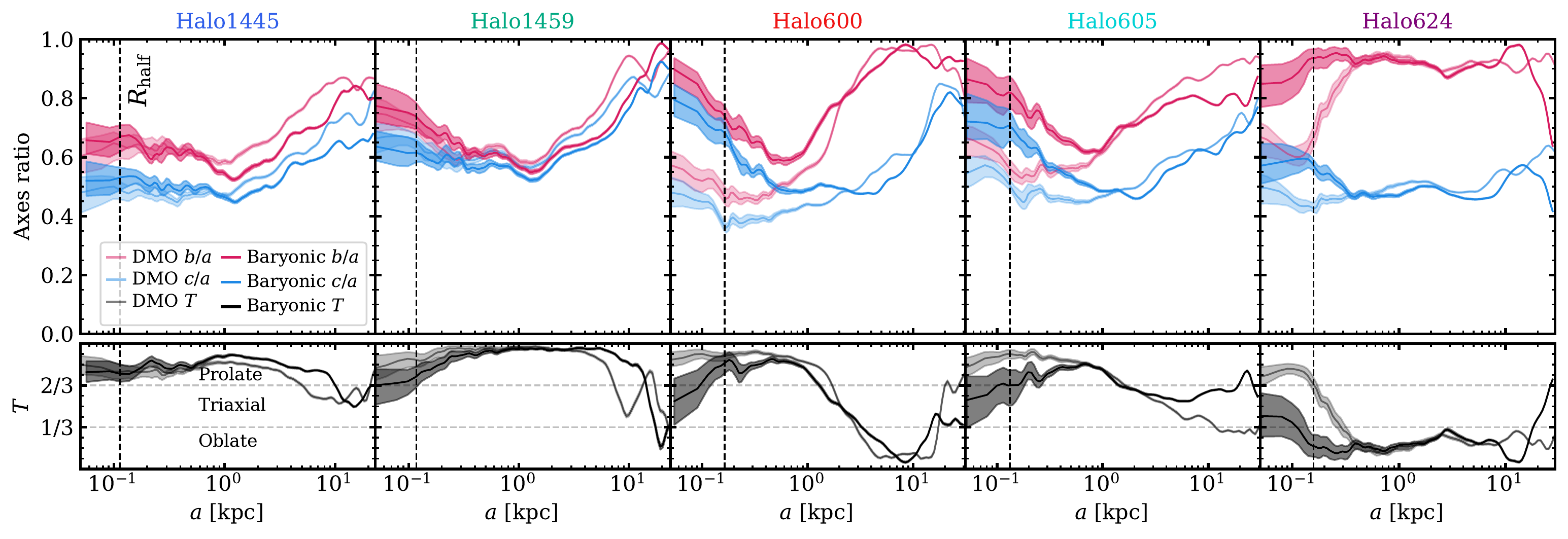}\\
\caption[Shape results for the \ac{EDGE} simulation suite at redshift zero.]{\textit{Upper panels:} The axial ratios $b/a$ (red) and $c/a$ (blue) for the \ac{DM} component in each of the \ac{EDGE} dwarf galaxies, shown over a radial range $50\,\rm{pc} \geqslant r \geqslant r_{\rm 200c}$.
\textit{Lower panels:} The triaxiality parameter $T$ (Equation \ref{equ:triaxiality}) over the same radial ranges.
\ac{DMO} simulations are shown with lighter lines, and baryonic physics with darker lines. A vertical dashed black line marks the stellar 3D half light radius in the baryonic simulation, averaged over the five most recent simulation outputs. The radial binning used for the shape algorithm has at minimum 5k particles per bin, and the result is smoothed with a Gaussian filter using $\sigma=1$. Only dwarf galaxies with $M_{200 \rm c} > 1.5\times10^9\,\rm{M}_{\odot}$ exhibit significant changes in their central halo shapes once baryonic physics are introduced.}
\label{fig:shape}
\end{figure*}

\section{Method} \label{method}

\subsection{Shape algorithm} \label{algorithm}

We estimate the \ac{DM} halo shape using the moment of inertia method described in \citealt{1991ApJ...368..325K,1991ApJ...378..496D,1992ApJ...399..405W}, and references therein. This fits homoeoidal shells with variable axis ratios and orientations to the \ac{DM} particle distribution. The advantages of using ellipsoidal shells as opposed to ellipsoidal volumes are explored in \citet{2011ApJS..197...30Z}, where it is found that thinner shells provide better fits to the \textit{local} particle distribution. However, these are more sensitive to deviations from ellipsoidal symmetry, especially due to substructure. We locate and remove all substructures in our \ac{EDGE} haloes using the \ac{AHF} \citep{2009ApJS..182..608K}. We define substructure as any bound subhalo consisting of 100 \ac{DM} particles or more. This substructure filtering dramatically improves both the accuracy and uncertainty of our shape fits over the regions that host substructure, and improves the legibility of our results. \par

For each halo, we divide the \ac{DM} distribution into a set of spherical shells that are spaced logarithmically from a maximum radius of $r_{\rm 200c}$ (where $r_{\rm 200c}$ is the radius of a sphere encompassing 200 times the critical density of the universe) to a minimum radius of $200\,$pc. We then transition to bins containing $5000$ initial particles for radii within $200\,$pc. This binning strategy is intended to ensure a reasonable minimum number of particles in every initial bin. \par

The shape and orientation of each shell is then adjusted, whilst keeping the major axis fixed, to fit the underlying particle distribution by solving the shape tensor $S$:
%
\begin{equation}
S_{ij} = \frac{\sum_{k}{m_{k} (\mathbf{r}_{k})_{i} (\mathbf{r}_{k})_{j}}} {\sum_{k}{m_{k}}},
\label{equ:shape}
\end{equation}
where the right-hand-side refers to the elements of $S$ in terms of a \ac{DM} particle $k$ with mass $m$ and a position vector $\mathbf{r}$. Axial ratios $a,b,c$ are derived from the eigenvalues of $S$ in the form $e_{x} = Mx^{2}/3$ (where $M$ is the total mass in the enclosed region and $x=a,b,c$). A rotation matrix for each shell is derived from the eigenvectors of $S$. This calculation is iterated, using the preceding shape as the new initial shell, until convergence criterion are met. We consider the shape fit to be converged when the difference between axial ratios vary by less than $10^{-4}$ between iterations. \citet{2022arXiv220911244F} show that the halo shape cannot be reliably determined within a flat density core. However, since none of the simulations analysed here are completely cored, this does not affect our analysis. Finally, we perform bootstrapping analysis on every shape fit, with $100$ iterations to approximate the $1\sigma$ uncertainty on both the axial ratios and orientation angle. We demonstrate the success of the algorithm in Appendix \ref{appendix:a}. \par

\subsection{Simulation suite}

The \ac{EDGE} project is a suite of fully cosmological zoom simulations produced with the \ac{AMR} tool {\sc ramses} \citep{2002A&A...385..337T}, focusing on isolated dwarf galaxies over the mass range $1.3\times10^9 \leqslant M_{\rm 200c}/\text{M}_{\odot} \leqslant 3.2\times10^9$. Our physics model produces galaxies with sizes and V-band brightnesses comparable to observed ultra-faint dwarf galaxies \citep[e.g.][]{2012AJ....144....4M, 2014MNRAS.439.1015K, 2019ARA&A..57..375S}. \ac{EDGE} assumes a \ac{LCDM} cosmology with cosmological parameters based on data from the Planck satellite \citep{2014A&A...571A..16P}, which are $\Omega_{\rm m}=0.309$, $\Omega_{\rm \Lambda}=0.691$, $\Omega_{\rm b}=0.045$ and $H_0=67.77\,\text{km\,s}^{-1}\,\text{Mpc}^{-1}$. Our setup and sub-grid physics are described fully in \citet{agertz2020}. \par

Of the full simulation suite, we analyse our highest resolution simulations (hydrodynamic grid resolution approaching $\sim 3$\,pc and $\sim 20\,\mathrm{M}_{\odot}$, \ac{DM} particle resolution of $\sim 120\,\mathrm{M}_{\odot}$). This is because tests performed in \S\ref{algorithm} indicate that an accurate shape fit requires particles of $\mathcal{O}(1000)$ or greater per radial bin, limiting the minimum radius that can be effectively probed at lower resolutions. We list all simulations analysed here in Table \ref{tab:edge_sims}, alongside some key properties. A more complete description of each simulation can be found in \citet{orkney}. \par

We perform a resolution study with lower resolution versions of each simulation in Appendix \ref{appendix:c}. This shows that, for radii over which the shape is well resolved, the axial ratios are reasonably converged to within $\Delta(x/a)=0.1$, where $x=b,c$. These small differences represent genuine changes in the \ac{DM} halo shape, rather than an inherent uncertainty in the algorithm, and arise due to tiny inadvertent changes in the assembly history and/or disruption of substructure. Therefore, we caution that shape comparisons between simulations with different resolutions and/or physics should do so with awareness of this convergence precision. \par

\begin{table*}
\centering
\begin{tabular}{l|ccc|l} 
\toprule
\toprule
\textbf{Name} & \textbf{$M_{200\rm c} (z=0)$} [M$_{\odot}$] & \textbf{$M_* (z=0)$} [M$_{\odot}$] & \textbf{$f_{\rm gas}$ ($z=0$)} & \textbf{Trivia} \\
\midrule
\textcolor{Halo1}{Halo1445} & $1.32\times10^9$ & $1.35\times10^5$ & $7.05\times10^{-6}$ & Reionisation fossil \\
\textcolor{Halo2}{Halo1459} & $1.43\times10^9$ & $3.77\times10^5$ & $3.11\times10^{-6}$ & Reionisation fossil \\
\textcolor{Halo5}{Halo600} & $3.23\times10^9$ & $9.84\times10^5$ & $7.09\times10^{-4}$ & Rejuvenator \\
\textcolor{Halo4}{Halo605} & $3.20\times10^9$ & $1.93\times10^6$ & $6.70\times10^{-4}$ & Rejuvenator \\
\textcolor{Halo3}{Halo624} & $2.65\times10^9$ & $1.08\times10^6$ & $5.23\times10^{-2}$ & Rejuvenator, $<1\,$kpc gas disc \\
\bottomrule
\bottomrule
\end{tabular}
\caption{A brief summary of the main dwarf galaxy simulations investigated in this paper. ``Reionisation fossil'' refers to galaxies that are quenched as a result of reionisation, whereas ``rejuvenator'' refers to galaxies that reignite star formation after this initial quenching. The gas fraction is given as $f_{\rm gas} = M_{\rm gas}/(M_{\rm gas} + M_{\rm DM})$ for all matter within the half light radius at $z=0$.}
\label{tab:edge_sims}
\end{table*}

\section{Results} \label{results}

\subsection{The halo shape} \label{shape}

We present the axial ratios of the \ac{DM} halo shape for each \ac{EDGE} dwarf as a function of the major-axis `$a$' in the upper panels of Figure \ref{fig:shape}, where we show the results for two physics schemes: a \ac{DMO} simulation and a baryonic simulation that includes gas and star formation. Axial ratios closer to unity, in particular the ratio $c/a$, indicate a more spheroidal configuration. In the lower panels, we quantify whether the haloes are prolate or oblate with the triaxiality parameter $T$ \citep{1991ApJ...383..112F}:
\begin{equation}
    T = \frac{1-b^{2}/a^{2}}{1-c^{2}/a^{2}},
    \label{equ:triaxiality}
\end{equation}
where $T>2/3$ is prolate, $T<1/3$ is oblate, and $1/3<T<2/3$ is triaxial (as labelled). \par

The two least massive haloes (Halo1445 and 1459) exhibit qualitatively comparable shape profiles for physics schemes, with differences mostly within $\Delta(x/a) = 0.1$, where $x=b,c$. The shape of each physics scheme deviates slightly at larger radii ($a\gtrapprox10\,$kpc), likely due to chaotic differences in their accretion histories. The halo at these larger radii is dominated by more recent accretion events, and is also less well phase-mixed than the inner halo, which leads to an increased stochasticity of the halo shape. \par

The three higher-mass haloes (Halo600, 605 and 624) exhibit comparable halo shape profiles for both physics schemes over the range $1 \gtrapprox a/\rm{kpc} \gtrapprox 10$, but the baryonic versions have raised axis ratios within $a=1\,$kpc. The change in the minor-to-major axis ratios exceed $\Delta(c/a)=0.2$ at the half light radius, which is greater than the usual standard deviation among halo populations \citep{2002ApJ...574..538J, 2005ApJ...629..781K, 2006MNRAS.367.1781A, 2007MNRAS.376..215B, 2019MNRAS.484..476C}. \par

The onset of this deviation begins well outside the half light radius in all three cases, suggesting that direct \ac{DM} heating from bursty star formation (which is limited to within the half light radius, see \citealp{orkney}) cannot be the main driver. These results may be highlighting a critical mass limit for which gas condensation is too mild to transform the primordial shape of a \ac{DM} halo. For our limited sample, this occurs somewhere above the mass of Halo1459 ($M_{\rm{200c}} = 1.43\times10^9$\,M$_\odot$; $M_{*} = 3.77\times10^5$\,M$_\odot$). A much larger statistical sample of simulations would be needed to refine this limit further, and also to account for the variability in halo assembly histories which can also affect the halo shape (see further discussion in \S\ref{GM}). This precise mass scale will be sensitive to our chosen sub-grid physics models, as it depends on the condensation of gas into the halo centre (which is itself dependent on the gas fraction) rather than the mass directly. We investigate this further in \S\ref{fblim}. \par

The \ac{DM} haloes of Halo1445, 1459, 600 and 605 all tend towards a slightly rounder shape in their outskirts ($a>10\,$kpc), with an increasingly triaxial halo configuration. This is consistent with the notion that the outer halo is constructed from more isotropic late-time accretion \citep[e.g.][]{2011MNRAS.416.1377V, 2023arXiv230208853C}. Halo1445 and Halo1459 are, across the full range of radii considered, more prolate than the higher-mass haloes. This suggests that these lighter haloes are constructed from more anisotropic accretion in the early universe. \par

\begin{figure}
\centering
  \setlength\tabcolsep{2pt}%
    \includegraphics[keepaspectratio, trim={0.3cm 0cm 0.3cm 0cm}, width=\columnwidth]{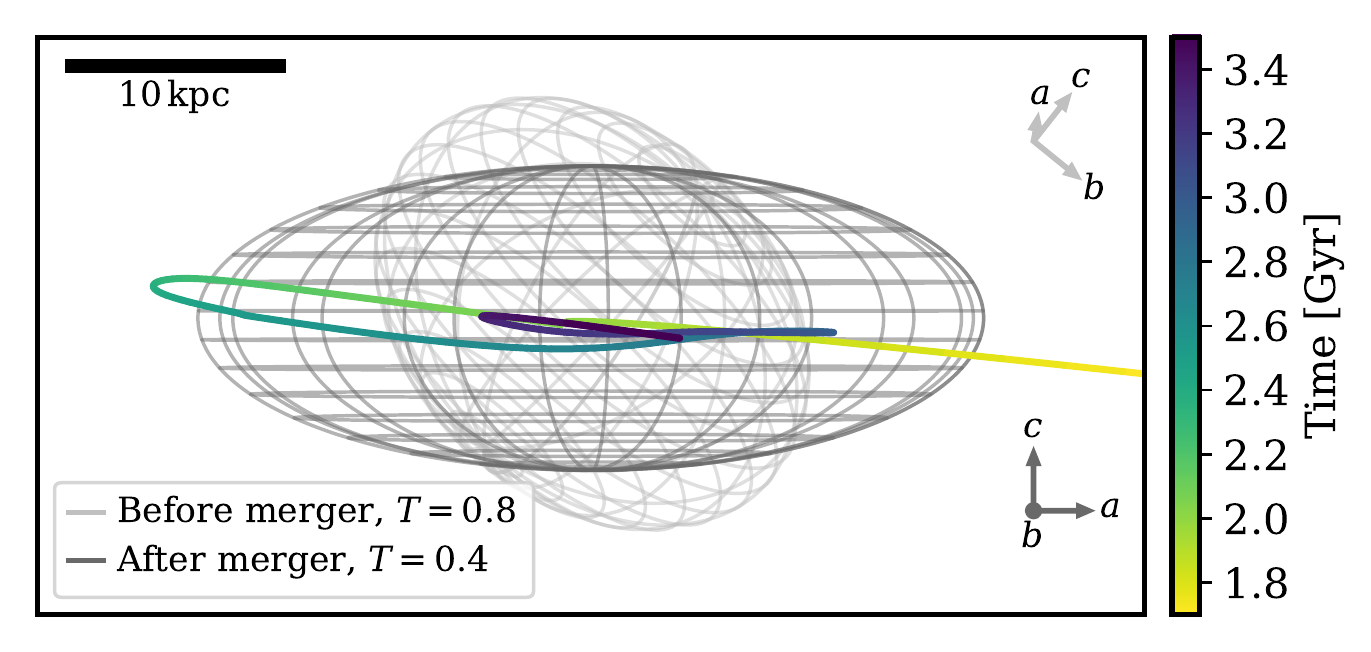}\\
\caption{A schematic illustrating the halo shape change resulting from a major merger in Halo624. The orbital path of the merger is represented by a coloured line, where the colour corresponds to the simulation time. The shape of \ac{DM} particles associated with the main progenitor halo, for radii $0.5\mathit{\textnormal{-}}10$\,kpc, is represented by an ellipsoidal wireframe (enlarged for clarity). The shape is highly prolate ($T=0.8$) before the merger event, and evolves to be more oblate ($T=0.4$) after the merger is complete. The frame has been oriented into the side-on plane of the final halo shape, which is well aligned with the side-on plane of the merger infall.}
\label{fig:shape_schematic}
\end{figure}

Halo624 is an outlier in that the halo shape is highly oblate beyond $a\sim500\,\rm{pc}$ for both baryonic and \ac{DMO} versions. This characteristic shape develops after a near equal-mass merger event (merger mass ratio $\sim1:1$), that can be seen in the merger tree just after $1.2\,$Gyr in Appendix \ref{appendix:b}, Figure \ref{fig:tree_624}. We investigate the repercussions of this merger event in Figure \ref{fig:shape_schematic}, which focuses on the halo shape in the range $0.5<a/\rm{kpc}<10$. This shows that an initially prolate halo ($T=0.8$) becomes significantly more oblate ($T=0.4$) after the merger remnant has decayed to the halo centre, and the new shape is well aligned with the plane of the merger infall. The shape transformation is caused by two compounding effects: i) the merging halo deposits a track of tidally-liberated \ac{DM} along the plane of infall, and ii) the strong gravitational influence of the merging object directly deforms the in-situ material. These combined effects are responsible for producing the oblate shape in both \ac{DMO} and baryonic versions of Halo624 at radii greater than $a\sim500\,$pc. This same merger is responsible for assembling a long-lived and rotating ex-situ stellar component, which manifests as a kinematically hot disc. \par

\subsection{The halo-gas alignment} \label{alignment}

\begin{figure}
\includegraphics[keepaspectratio, trim={0.3cm 0.5cm 0.2cm 0cm}, width=\columnwidth]{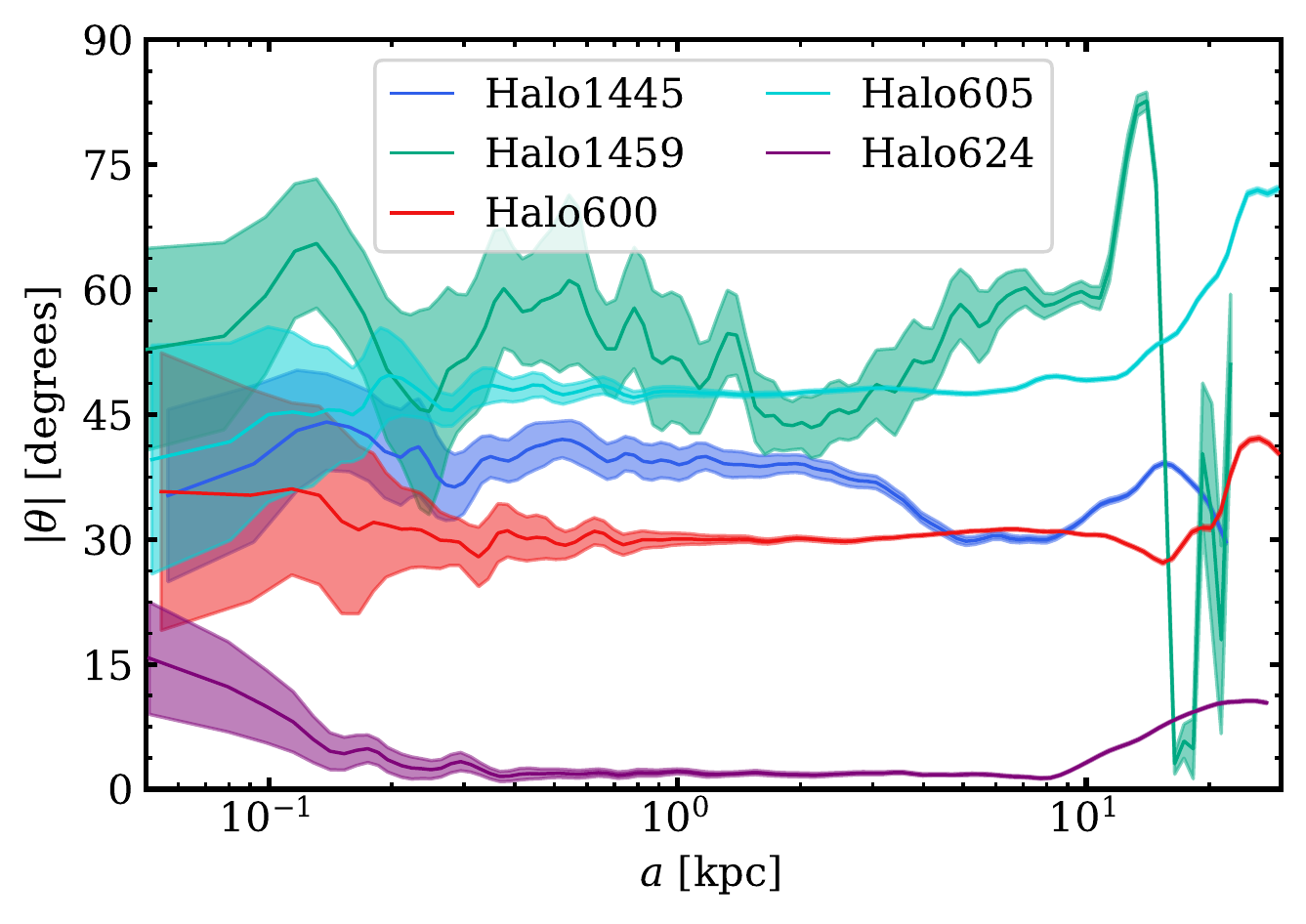}\\
\caption{The absolute alignment angle $\theta$ between the short-axis $c$ and the angular momentum vector of the cool central gas ($<1\,$kpc, $<10^4$\,K where available). An angle of $\theta=0$ corresponds to perfect alignment. The shaded regions indicate the $1\sigma$ bootstrap uncertainty. Larger uncertainties reflect bins with fewer \ac{DM} particles, and/or regions that are highly spheroidal. Only Halo624 has a well-defined rotating gas disc, and in this case the gas is well aligned with the \ac{DM} halo. }
\label{fig:alignment}
\end{figure}

Stellar and gas discs tend to align with the shape and rotation of the surrounding \ac{DM} halo \citep[e.g.][]{2008ApJ...681.1076D}. The cosmic filaments that feed the \ac{DM} growth also feed the baryonic growth, and so it is natural that they should produce similar configurations. \par

In Figure \ref{fig:alignment}, we present the absolute alignment angle $\theta$ between the \ac{DM} minor axis, $c$, and the angular momentum vector of the central cool gas. The central cool gas is defined as all gas with $T\leqslant1\times10^4\,\text{K}$ within 1\,kpc of the halo centre, or the coolest 10 per cent of gas within 1\,kpc of the halo centre if no $<1\times10^4$\,K gas is available. An angle of $\theta=0$ corresponds to perfect alignment between halo shape and gas angular momentum. \par

The results show that, for all haloes, the orientation of the \ac{DM} is relatively consistent out to a radius of $a\sim10\,$kpc, at which point it begins to deviate. This reflects how the \ac{DM} halo within $a\sim10\,$kpc is constructed primarily from thin filamentary accretion, whereas the halo outskirts experience increasingly isotropic accretion from thicker filaments. This behaviour is particularly evident in Halo1459. \par

Halo624, which is the only halo to possess a meaningful (albeit small) rotating gas disc (see the visualisation in Figure 2 of \citealt{orkney}), has exceedingly good alignment. The other \ac{EDGE} dwarfs lack any coherent gas structure, and show no signs of alignment between the angular momentum of their gas and the halo shape. \par

\subsection{Central shape evolution} \label{evolution}

\begin{figure*}
\includegraphics[keepaspectratio, trim={0.5cm 0cm 1cm 0cm}, width=\linewidth]{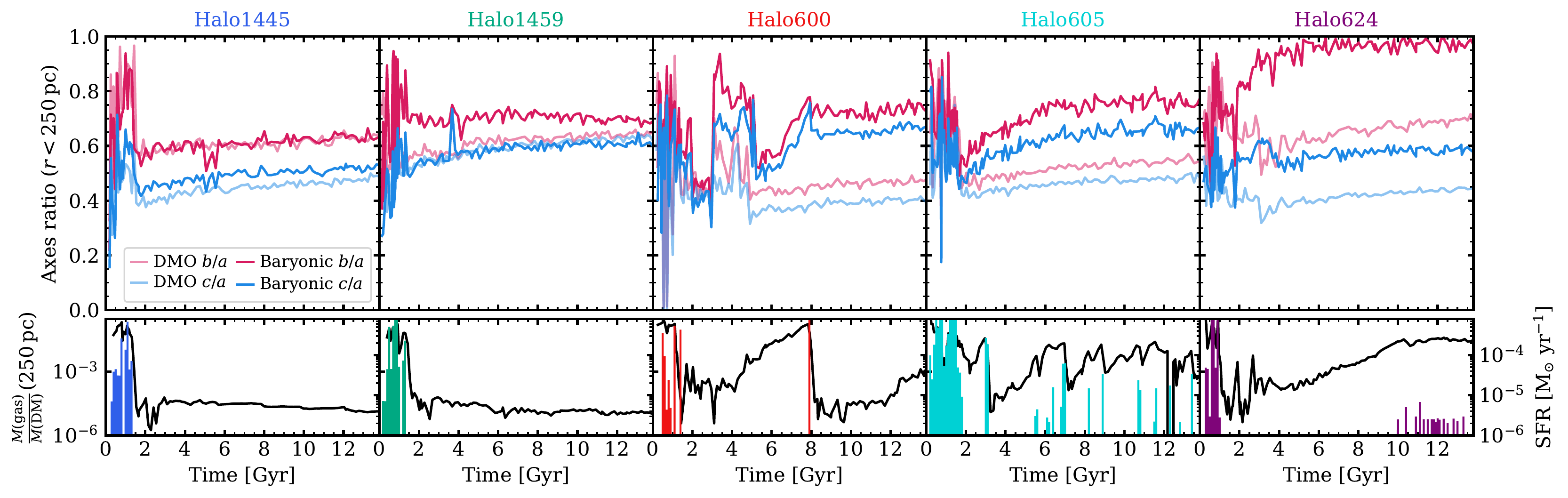}\\
\caption[The central shape evolution since redshift four for the \ac{EDGE} simulation suite.]{\textit{Upper panels:} The axial ratios $b/a$ (red) and $c/a$ (blue) for the central ($r<250\,\rm{pc}$) \ac{DM} component in each of the \ac{EDGE} dwarf galaxies with time. \ac{DMO} simulations are shown with lighter lines, and baryonic physics with darker lines.
\textit{Lower panels:} The ratio of gas to \ac{DM} mass at 250\,pc (black line, left axis) and the star formation history of the main progenitor halo in 100\,Myr bins (coloured histogram, right axis) with time.
Changes in the central halo shape are correlated with periods of active star formation and central gas mass fluctuations.}
\label{fig:shape_with_time}
\end{figure*}

The time evolution of the central halo is presented in Figure \ref{fig:shape_with_time}. We show the evolution of the mean central ($0 < r/\rm{pc} < 250$, where $r$ is the radius of the initial spherical bin) \ac{DM} halo shape in the upper panels. We show the gas to \ac{DM} mass ratio at $250$\,pc (left axis), alongside the star formation history (right axis), in the lower panels. The halo shape is extremely stochastic during the rapid assembly period in the first 2 Gyrs, and so it is difficult to decipher its evolution. After this early assembly, all haloes display a sub-dominant and near-linear growth in their axial ratios. This is likely capturing the constant shape transformation due to the random deflection of \ac{DM} particle orbits, which persists at a low level throughout all times regardless of any baryonic presence. \par

For the lighter haloes (Halo1445 and Halo1459), the shape evolution is relatively featureless after the initial assembly period. The gas to \ac{DM} mass ratio of the baryonic versions are extremely low ($<10^{-4}$) with no star formation after 2\,Gyr. This implies that there is insufficient gas condensation to further perturb \ac{DM} particle orbits, and so the shape of both physics schemes remain roughly equivalent. \par

The more massive haloes all achieve a rejuvenation in their star formation after initially quenching. In \citet{orkney}, we showed that the gravitational potential fluctuations driven by bursty star formation have a minimal effect on the central density of the \ac{DM}. But the re-condensation of material from the gas flows that they launch may still prove to affect the inner halo shape. \par

Halo600 exhibits a sudden increase in its axial ratios at $\sim3\,$Gyr in both physics versions. This is the result of multiple massive merger events occurring in quick succession, which can be seen from inspection of the merger tree in Appendix \ref{appendix:b}, Figure \ref{fig:tree_600}. These merger events disrupt the \ac{DM} halo and warp its shape, and the halo does not fully settle until $\sim 5\,$Gyr. The mass provided by these mergers deepens the potential well and draws more gas into the halo centre, which ultimately triggers a renewed burst of star formation at 8\,Gyr. The baryonic Halo600 evolves towards a rounder shape between 5\,Gyr and 8\,Gyr ($\Delta(c/a)>0.2$), which is correlated with the extended period of gas condensation. This shape evolution is halted by the massive gas blowout driven by \ac{SNe} feedback at 8\,Gyr. \par
The baryonic Halo605 evolves towards raised axial ratios over a simulation time of $\sim2\mathit{\textnormal{-}}7\,$Gyr. This behaviour is once again correlated with a massive gas outflow and inflow resulting from a sudden \ac{SNe} blowout and gas re-condensation. \par

The baryonic Halo624 adopts a highly oblate central shape from $\sim2\,$Gyr onwards, in significant contrast with its \ac{DMO} counterpart. This sudden change is well-correlated with the near equal-mass early merger event that is discussed in \S\ref{shape}. This difference can be attributed to two effects. Firstly, the \ac{DMO} version of Halo624 has a cuspy host density profile, whereas the baryonic version is slightly flattened from $\sim1\,$Gyr onward due to baryonic heating effects \citep[see][]{orkney}. The dense halo in the \ac{DMO} version helps stave off tidal interference from the merger, and the merger is itself dissolved by gravitational tides before its orbit reaches the inner radii. Conversely, the merger in the baryonic version is able to penetrate to these inner radii before it is completely dissolved. Secondly, the merger in the baryonic version contains a peak gas mass of $7\times10^6\,\rm{M}_{\odot}$, roughly half that in the main progenitor at the same time. The shocks from this gas accretion contribute to the large-scale fluctuations in the central gas mass ratio around $2\,$Gyr (see Figure \ref{fig:shape_with_time}), which in turn contribute to the drastic inner shape transformation. Halo624 rejuvenates its star formation in the baryonic version, but the resulting stellar feedback is not capable of driving large-scale gas flows. \par

\subsection{Velocity anisotropy} \label{anisotropy}

\begin{figure*}
\includegraphics[keepaspectratio, trim={0.3cm 0cm 0.2cm 0cm}, width=\linewidth]{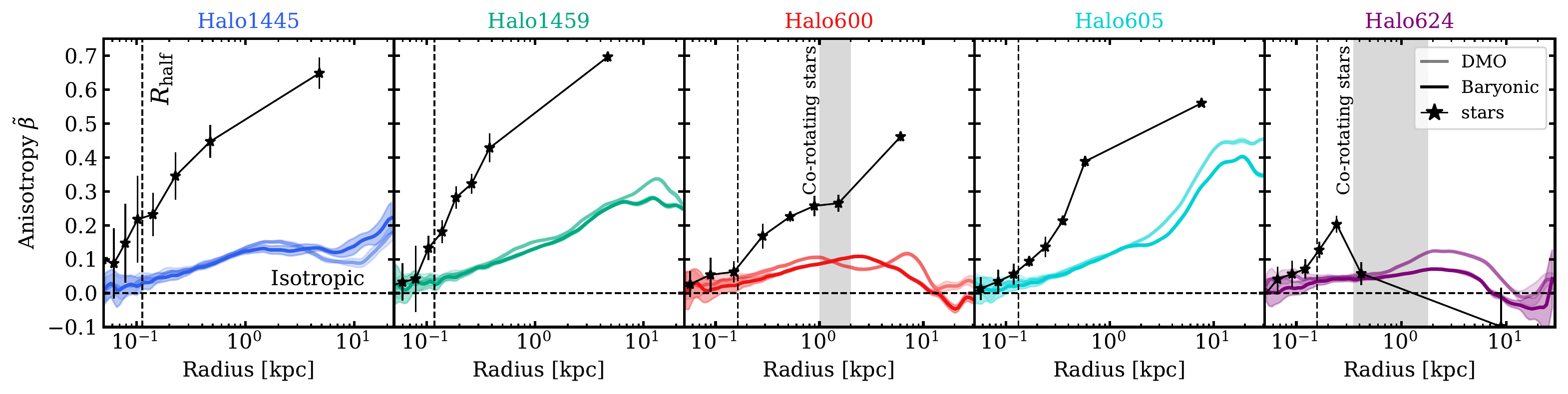}\\
\caption[The velocity anisotropy for the \ac{EDGE} dwarfs.]{The symmetrized velocity anisotropy parameter $\tilde{\beta}$ over the range 50\,pc to $r_{\rm 200c}$ for the stars (black lines) and \ac{DM} (coloured lines) for the baryonic (darker lines) and \ac{DMO} (lighter lines) \ac{EDGE} simulations. The results shown are the median over the penultimate ten simulation outputs ($0.1 \leqslant z \leqslant 0$). The shaded regions and error bars indicate the standard deviation over those ten outputs, demonstrating the reduced certainty at the lowest and highest radii. A black dashed line marks $\tilde{\beta}=0$, which represents isotropy. A vertical dashed black line marks the stellar 3D half light radius in the baryonic simulation, averaged over the five most recent simulation outputs. Results are smoothed with a Gaussian filter using $\sigma=1$. Halo600 and Halo624 both possess co-rotating stellar components over a specific range of radii, and these are indicated with grey shaded bands.}
\label{fig:anisotropy}
\end{figure*}


The velocity anisotropy parameter gives a measure of the how tangentially biased (i.e. more circular), or how radially biased (i.e. more plunging) the orbital distribution is \citep{1980MNRAS.190..873B, 2008gady.book.....B}. Here we use a symmetrized version as in \citet{2017MNRAS.471.4541R}, which constrains the lower limit to $-1$ as opposed to $-\infty$:
\begin{equation}
\tilde{\beta} = \frac{\sigma_{\rm r}^2 - \sigma_{\rm t}^2}{\sigma_{\rm r}^2 + \sigma_{\rm t}^2} = \frac{\beta}{2-\beta},
\label{equ:betas}
\end{equation} 
where $\sigma_t$ is the transverse velocity dispersion and $\sigma_r$ is the radial velocity dispersion. $\tilde{\beta}=-1$ corresponds to fully tangential orbits, $\tilde{\beta}=0$ corresponds to fully isotropic orbits and $\tilde{\beta}=1$ corresponds to fully radial orbits. \par

Previous works have shown that \ac{DM} haloes are described by near-isotropic central regions that grow more radially biased with radius (with a characteristic $\tilde{\beta}\approx0.2$), and then either plateau or return to isotropy towards the virial radius \citep{2010MNRAS.406..922T, 2011MNRAS.415.3895L, 2012ApJ...752..141L, 2012JCAP...10..049S, 2021MNRAS.501.5679C}. \citet{1997MNRAS.286..865T} associate this trend with the recent accretion of substructure and \ac{DM} from the cosmic web, which leaves the outskirts of the halo in a dynamically unrelaxed state (see also \citealt{2013MNRAS.430..121P}). A more tangential $\tilde{\beta}$ at the halo outskirts can be caused by an injection of angular momentum from near-equal mass mergers \citep{2004MNRAS.354..522M, 2007MNRAS.376.1261M, 2012JCAP...10..049S}, with the effect being relatively insensitive to the merger impact parameter \citep{2007MNRAS.376.1261M}. In this sense, both halo shape and velocity anisotropy are heavily influenced by the particular assembly history. \par

We present radial $\tilde{\beta}$ profiles for each \ac{EDGE} dwarf in Figure \ref{fig:anisotropy}, for the stars (black lines) and \ac{DM} (coloured lines), for both the baryonic (darker lines) and \ac{DMO} (lighter lines) simulations. This calculation is performed in spherical bins, consistent with the definition of $\tilde{\beta}$. We take the median of the results over the penultimate ten simulation outputs ($0.1 \leqslant z \leqslant 0$) to account for any temporal fluctuations. The individual $\tilde{\beta}$ profiles over this timeframe show little evolution, and we illustrate this by including the $1\sigma$ standard deviation as shaded bands. \par

\subsubsection{Dark matter anisotropy}

The $\tilde{\beta}$ profiles are almost identical between both physics schemes. The baryonic versions of our more massive haloes (Halo600, Halo605 and Halo624) are slightly more isotropic in their central regions than their \ac{DMO} counterparts, but this difference is within the temporal variation -- and the \ac{DMO} simulations are already highly isotropic at their centres. \citet{2019MNRAS.484..476C} show that there is a greater discrepancy between \ac{DMO} and baryonic physics at higher mass scales than in \ac{EDGE}, with baryonic simulations adopting more isotropic distributions. \par

Halo1445, Halo1459 and Halo605 all exhibit trends of increasing $\tilde{\beta}$ with radius, as expected for haloes where the outer regions are assembled via mergers with radially biased infall trajectories. The $\tilde{\beta}$ profiles of Halo600 and Halo624 are, in comparison, more tangential for radii $\gtrapprox 1$\,kpc. These two haloes are also comparatively more oblate for radii $\gtrapprox 1$\,kpc (see Figure \ref{fig:shape}). A common process, possibly involving the angular momentum of accreted material, could be responsible for both properties. Both Halo600 and Halo624 assemble from the merging of near equal-mass progenitors at high redshift, as opposed to the other \ac{EDGE} haloes which experience more minor accretion events. The isotropy at these radii is set during this relatively early epoch. For context, see the merger tree visualisations presented in Appendix \ref{appendix:b}. \par

\subsubsection{Stellar anisotropy} \label{stars_DM}

\begin{figure*}
\includegraphics[keepaspectratio, trim={0.3cm 0cm 0.2cm 0cm}, width=\linewidth]{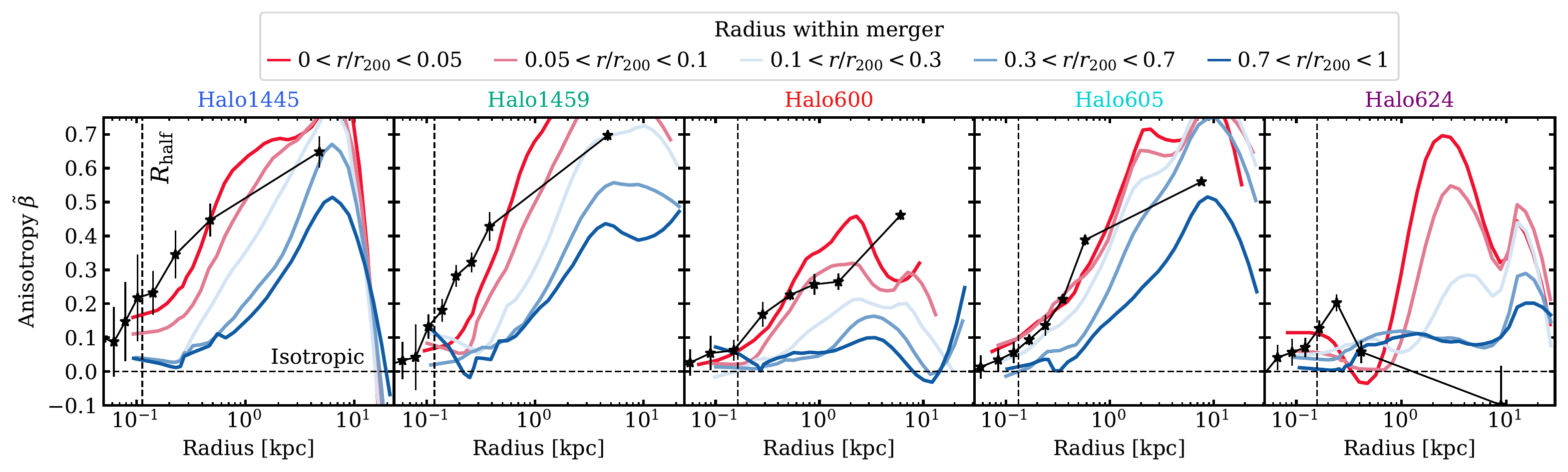}\\
\caption{The symmetrized velocity anisotropy parameter $\tilde{\beta}$ for the baryonic simulations, but only for the \ac{DM} that originated from the most major mergers (merger mass ratio $>1:30$) after $z=6$. The \ac{DM} is partitioned according to its orbital radius within the merging halo at the snapshot prior to infall. The stellar anisotropy profiles from Figure \ref{anisotropy} are overlayed, which illustrates that the \ac{DM} from the centres of major mergers share a similarly raised anisotropy. The \ac{DM} anisotropy profiles have been smoothed with a Savgol filter.}
\label{fig:merger_anisotropy}
\end{figure*}

The stellar distribution of each \ac{EDGE} galaxy includes a dense in-situ central component and a wider ex-situ component that rapidly diminishes towards higher radii. The $\tilde{\beta}$ for the stars approaches $\tilde{\beta}=0$ at low radii, and then climbs towards more radial orbits with increasing radius. The $\tilde{\beta}$ profile slope is higher for the stellar distribution than the \ac{DM}, typically exceeding $\tilde{\beta}=0.5$ for $r\gtrsim 3 R_{\rm half}$. The exception to this is Halo624, which is the only simulated \ac{EDGE} dwarf, presented in this paper, that hosts a coherent stellar and gas disc. For this dwarf, the velocity anisotropy becomes more tangential beyond $r\gtrsim 2 R_{\rm half}$. This is due to the co-rotating stellar component, which has a net rotational velocity of $9\,\text{km}\,\text{s}^{-1}$, and depresses the velocity anisotropy over the radii that it is active ($0.35<r/\rm{kpc}<1.8$). Similarly, there is a vague co-rotating feature in Halo600 with a net rotational velocity of $7\,\text{km}\,\text{s}^{-1}$ over the radii $1<r/\rm{kpc}<2$, though this appears to have only a marginal impact on $\tilde{\beta}$ over the same radii. Both of these features owe to the historic assembly of the galaxy, rather than being born from any native rotating gaseous component. \par

We suggest that this is primarily due to two factors. Firstly, the fraction of stars formed ex-situ increases rapidly with radius (see \citealt{2023arXiv230705130G}). These stars are accreted from only the most massive mergers because lighter mergers do not reach the conditions necessary for star formation, and generally infall along more radial trajectories due to the effects of dynamical friction \citep[as in][]{2022ApJ...926..203V}. Secondly, these stars are tightly bound at the centres of each merging halo, and are thereby shielded from tidal disruption for an extended period of time \citep[similar to the effects in][]{2021arXiv211201265V}. The orbit of the merger remnant retains a highly radial velocity, and the galacocentric distance decreases due to dynamical friction. The stars then inherit these properties when the remnant is fully disrupted, which is predisposed to occur near the galactic centre where the gravitational tides are strongest. Following this argument, the \ac{DM} from the centres of these merging haloes should share a similar radial anisotropy to that of the stellar component. \par

To test this hypothesis, we identify the most major mergers for each \ac{EDGE} simulation and tag the \ac{DM} particles within their pre-infall virial radius. We show the final velocity anisotropy of this \ac{DM} in Figure \ref{fig:merger_anisotropy}, where the \ac{DM} is partitioned according to its orbital radius within the pre-infall merging halo. The most loosely-bound \ac{DM} ($0.7<r/r_{\rm 200c}<1$) tends to have a slightly more radial $\tilde{\beta}$ than the full \ac{DM} distributions in Figure \ref{anisotropy}, reflecting the more radial infall trajectory of the host merger. The $\tilde{\beta}$ slope becomes increasingly radial as we consider \ac{DM} from deeper within the merging haloes (lighter lines). The innermost radial bin ($0<r/r_{\rm 200c}<0.05$) has a slope comparable to the stellar profiles, supporting our prediction. \par

There are also physical reasons that the in-situ stellar component may possess a radially velocity bias. Some stars will have inherited the velocity of the infalling gas clumps from which they formed, or otherwise been scattered onto wider orbits following an encounter with a merger \citep[e.g.][]{2009ApJ...702.1058Z, 2015MNRAS.454.3185C}. \par

\subsection{Accretion anisotropy} \label{a_anisotropy}

\begin{figure}
\includegraphics[keepaspectratio, trim={0.3cm 0.5cm 0.3cm 0cm}, width=\columnwidth]{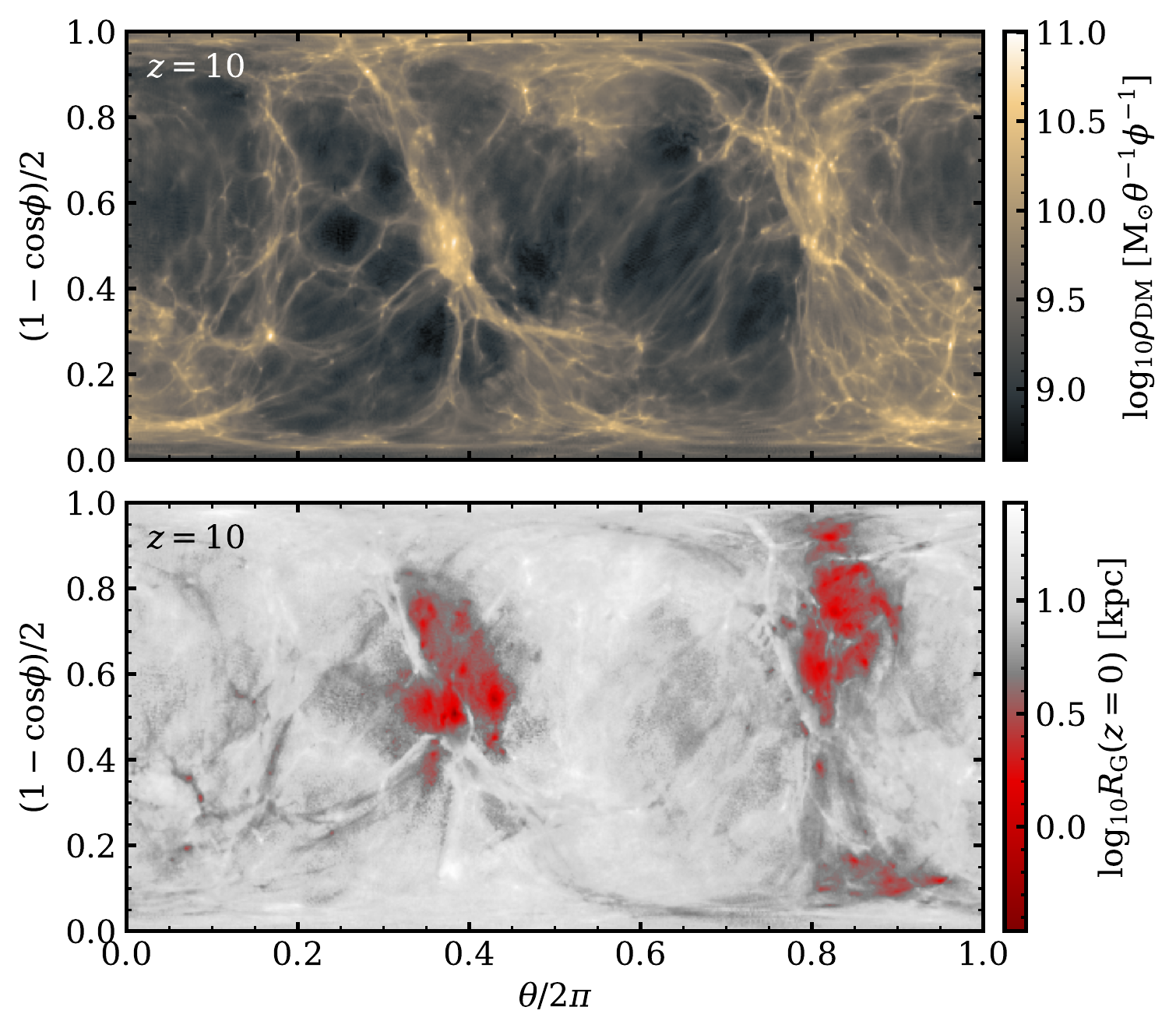}\\
\caption[]{Spherical polar coordinate maps of the \ac{DM} material within the virial radius of Halo605 at $z=0$, shown at its location at $z=10$. The upper panel shows the angular mass density, whereas the lower panel shows the final orbital radius within the host halo at $z=0$. The \ac{DM} within the inner few kpc at $z=0$ was accreted from a narrow range of angular directions.}
\label{fig:anglemap}
\end{figure}

Here, we briefly consider the angular anisotropy of accreted \ac{DM}. In Figure \ref{fig:anglemap}, we show the \ac{DM} within the virial radius of Halo605 at $z=0$, but as it appeared at $z=10$. The \ac{DM} is shown in spherical polar coordinate angles, which are normalised to uniformity within the range $0\mathit{\textnormal{-}}1$, with the origin taken as the position of the main progenitor at $z=10$. The upper panel shows the mass density, revealing a complex web of filaments and nodes. The lower panel shows the final galactocentric radius of the material within Halo605 at $z=0$. This illustrates that the halo material within $R_{\rm G}=1\,$kpc originated from a few key angular directions, whereas the outer halo is constructed from more isotropically distributed material (and whilst not shown here, this material remains more isotropically distributed even as it coalesces under gravity). The other \ac{EDGE} galaxies reveal comparable patterns. \par

A more detailed investigation into the accretion isotropy in \ac{LCDM} haloes at higher masses is performed in \citet{2011MNRAS.416.1377V, 2023arXiv230208853C}, and affirms our interpretation here: later assembly is characterised by more isotropic accretion, which yields a rounder \ac{DM} distribution. See also \citet{1997MNRAS.290..411T, 2018MNRAS.476.1796S, 2018MNRAS.481..414G}. \par

\subsection{Altered assembly histories with genetic modification} \label{GM}

\begin{figure}
\centering
  \setlength\tabcolsep{2pt}%
    \includegraphics[keepaspectratio, trim={0.3cm 0cm 0.2cm 0cm}, width=\columnwidth]{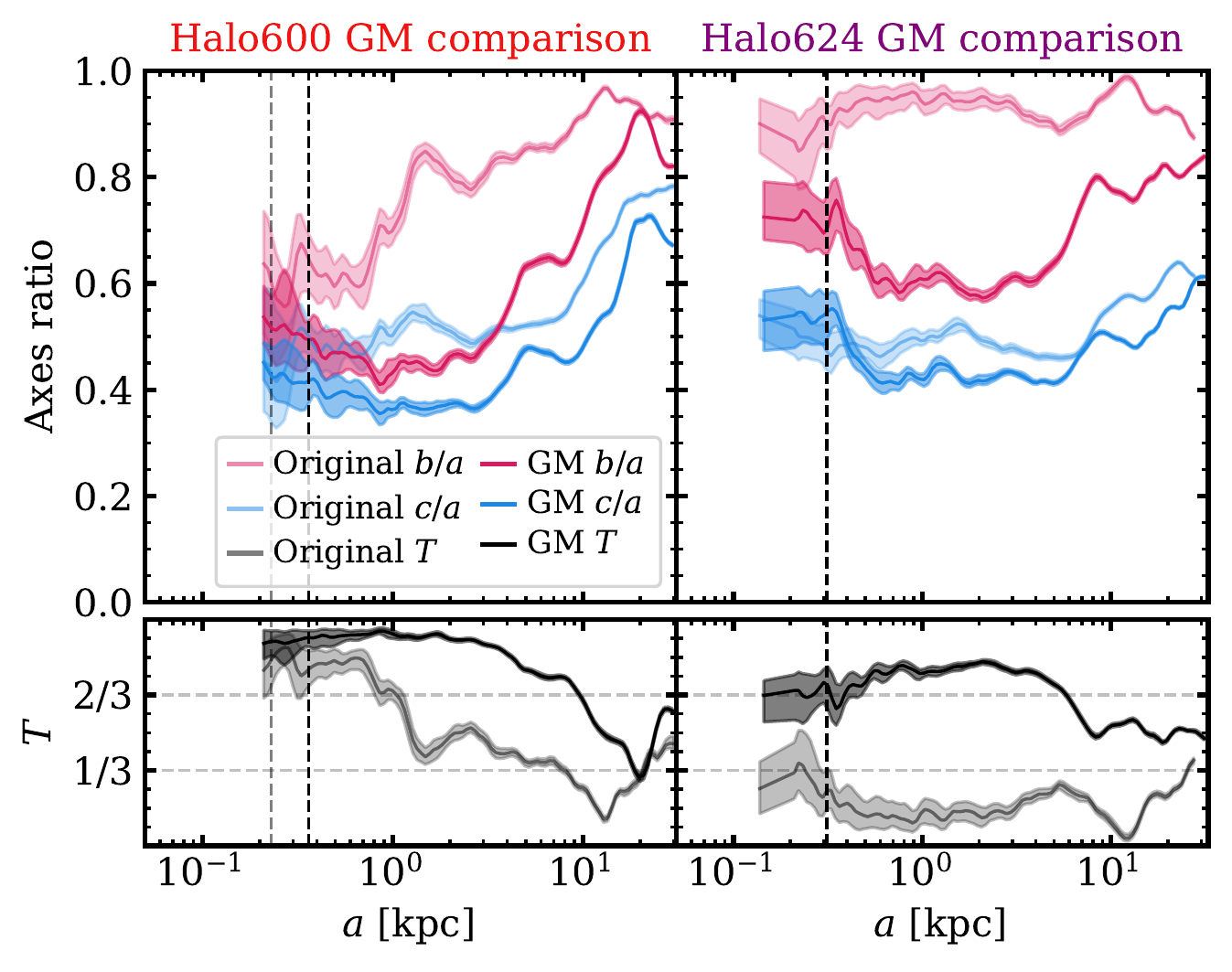}\\
\caption{The same form as Figure \ref{fig:shape}, but now comparing the original simulations (light lines) to genetically modified simulations with a different mass assembly (dark lines). Both simulations are run at the lower `fiducial' resolution, which limits the minimum effective radius of the shape fit. Changing the accretion history can have a dramatic impact on the halo shape at all radii.}
\label{fig:GM_shape}
\end{figure}

\begin{figure}
\centering
  \setlength\tabcolsep{2pt}%
    \includegraphics[keepaspectratio, trim={0.3cm 0cm 0.2cm 0cm}, width=\columnwidth]{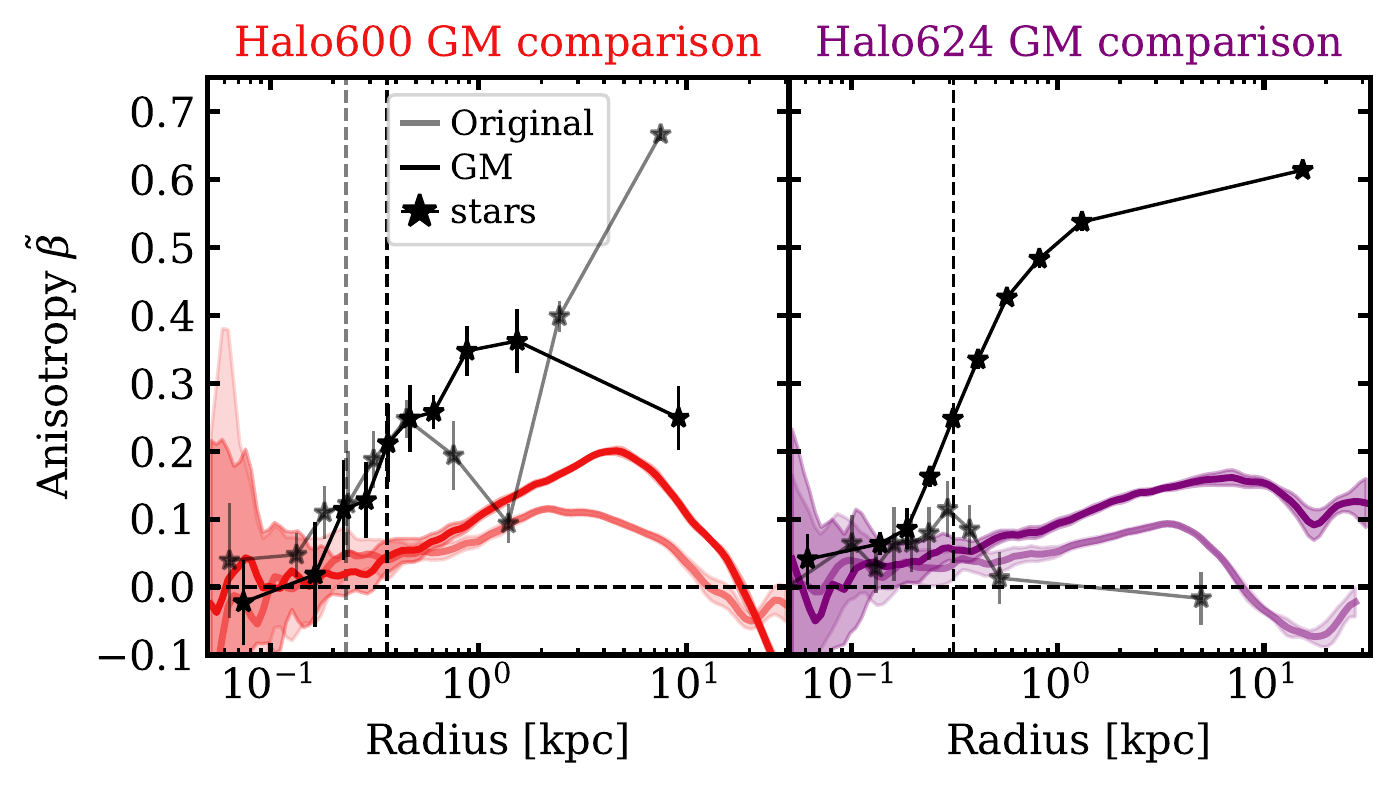}\\
\caption{The same form as Figure \ref{fig:anisotropy}, but now comparing original simulations (light lines) to genetically modified simulations (dark lines). Both simulations are run at the lower `fiducial' resolution. Changing the accretion history has huge implications for the velocity anisotropy of both stars and \ac{DM} beyond the half light radius.}
\label{fig:GM_anisotropy}
\end{figure}

The \ac{EDGE} simulation suite incorporates genetically modified (GM) initial conditions \citep[see][]{2016MNRAS.455..974R, 2018MNRAS.474...45R, genetic}. This allows for the accretion history of a target halo to be modified in a controlled manner. This approach has already supported investigations into the assembly of ultra-faint dwarfs \citep{rey2019}, the rejuvenation of quenched dwarf galaxies \citep{rey2020}, and more (e.g. \citealp{2022arXiv221115689R, 2023arXiv230705130G}). \par

Thus far, our results have indicated that the shape and velocity anisotropy in dwarf \ac{DM} haloes is highly dependent on the way in which that halo was constructed. This is especially true for Halo600 and Halo624, whose early assembly is dominated by major merger events. The \ac{DM} halo of Halo600 is oblate and has a more isotropic velocity distribution from $\sim2\,$kpc and outwards. We link these properties to the rapid halo assembly that begins from $\sim3\,$Gyr, where several similar-mass haloes coalesce in a high angular momentum merger event. The \ac{DM} halo of Halo624 is oblate outside of the half light radius, and also exhibits more isotropic velocity distributions in both stars and \ac{DM} over these same radii. We link these properties to an early major merger, which we illustrated in Figure \ref{fig:shape_schematic}. \par

In order to scrutinize these narratives, we compare with a GM version of each dwarf, hereafter Halo600 GM and Halo624 GM. Halo600 GM has a delayed assembly, and the merger frequency is smoothed over a broader time period (see Appendix, \ref{appendix:b} Figure \ref{fig:tree_600_GM}). Halo624 GM has a greater final halo mass, and several of its massive mergers occur at an earlier time (see Appendix \ref{appendix:b}, Figure \ref{fig:tree_624_GM}). A full description of these modifications, and the methods by which they were designed, is included in \citealt{rey2020}. The GM dwarfs are simulated at our fiducial resolution of $m_{\rm DM}=960\,\rm{M}_{\odot}$, as opposed to the high resolution of $m_{\rm DM}=120\,\rm{M}_{\odot}$ that is used elsewhere in this work. We stress that these limited resolutions are less effective in probing the inner halo shape transformations that are discussed in Sections \S\ref{shape} and \S\ref{evolution}, both due to their raised minimum resolved radius and the quickened disruption of infalling clumps, therefore we avoid interpreting the simulations in this respect. Nonetheless, the shape properties remain reasonably well resolved for radii $>0.2\,$kpc, and more generally we find that the shape and velocity anisotropy profile properties are superficially insensitive to this change in mass resolution. We confirm this with a resolution comparison in Appendix \ref{appendix:c}. Throughout this section, we will compare the GM simulations to their corresponding versions at this same lower fiducial resolution. \par

We compare the \ac{DM} halo shape between the original and GM versions in Figure \ref{fig:GM_shape}. Halo600 GM is far more prolate than the unmodified version, but still returns to an oblate shape at the virial radius. This is a consequence of the delayed assembly: the halo within $a\approx10\,$kpc is constructed more gradually in the GM version, leading to a retention of the primordial prolate shape. Then, the delayed major mergers contribute to an oblate shape at the halo outskirts. In effect, the impact of the mergers on the halo triaxiality profile has been shifted towards higher radii. Halo624 GM is more prolate over all radii, in stark contrast with the original simulation. The early major merger shown in Figure \ref{fig:shape_schematic} still occurs, but its effects are completely overwhelmed by other major mergers which now occur at earlier times and therefore impact the inner halo more considerably. \par

We compare the velocity anisotropy profiles in Figure \ref{fig:GM_anisotropy}. The \ac{DM} $\tilde{\beta}$ in both GM versions is more radial from $\approx1\,$kpc and outwards. There is a seeming correlation between a radial velocity anisotropy and a prolate halo shape. Halo600 GM no longer possesses any co-rotating stellar features, and accordingly the stellar anisotropy is more radial over the range $1<r/r_{\rm 200c}<2$. In Halo624 GM, the anisotropy profiles are once again more radial. Again, the co-rotating stellar feature has vanished. The stellar profile is instead dominated by stars from the last major merger, which fell in along a steep radial trajectory. These stars have subsequently adopted an extremely radial $\tilde{\beta}$, in agreement with our expectations from \S\ref{stars_DM}. \par

\subsection{Weak feedback models} \label{fblim}

\begin{figure}
\centering
  \setlength\tabcolsep{2pt}%
    \includegraphics[keepaspectratio, trim={0.3cm 0cm 0.2cm 0cm}, width=\columnwidth]{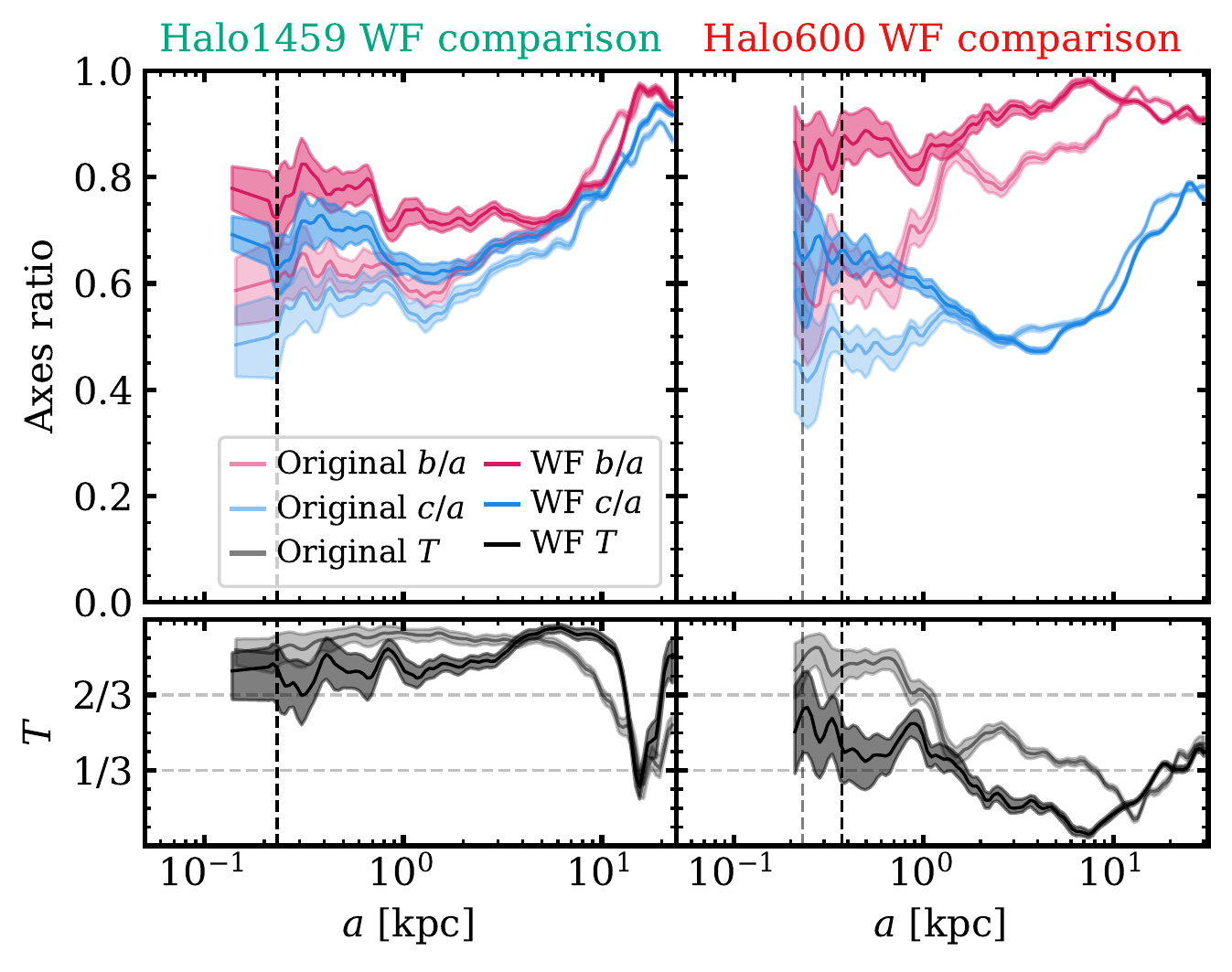}\\
\caption{The same form as Figure \ref{fig:shape}, but now comparing original baryonic dwarfs (light lines) to those simulated with a weak feedback model (`WF', dark lines), which places an upper limit on the temperature and velocity of stellar \ac{SNe} feedback. Both simulations are run at the lower `fiducial' resolution, which limits the minimum effective radius of the shape fit. Less feedback has supported more star formation and higher gas densities, leading to more substantial shape transformations than in the original versions.}
\label{fig:fblim_shape}
\end{figure}

\begin{figure}
\centering
  \setlength\tabcolsep{2pt}%
    \includegraphics[keepaspectratio, trim={0.3cm 0cm 0.2cm 0cm}, width=\columnwidth]{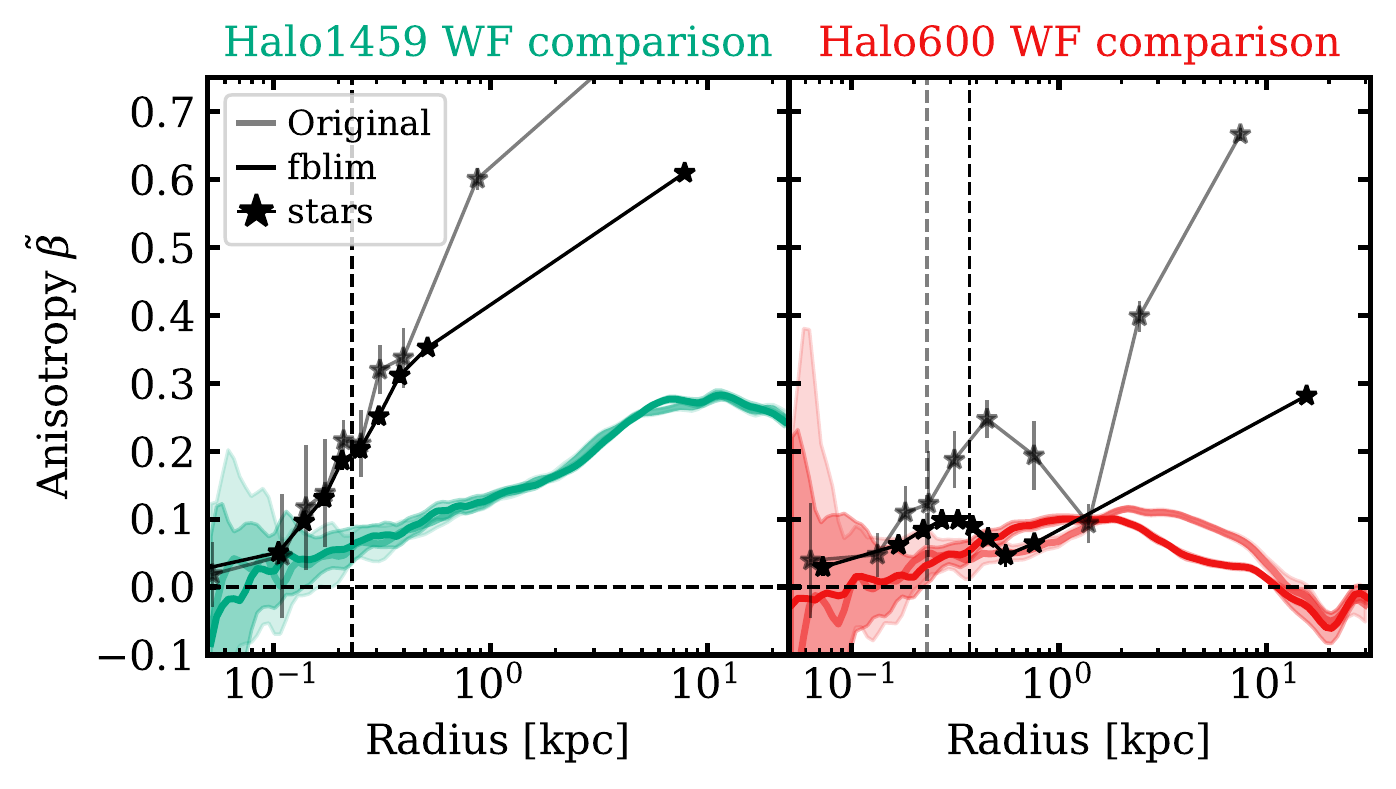}\\
\caption{The same form as Figure \ref{fig:anisotropy}, but now comparing original simulations (light lines) to those simulated with a weak feedback model (`WF', dark lines). Both simulations are run at the lower `fiducial' resolution. The stellar anisotropy profiles are affected by the feedback physics, but the \ac{DM} is relatively unchanged.}
\label{fig:fblim_ansitropy}
\end{figure}

In \S\ref{evolution}, we associated the transformation of the \ac{DM} halo shape within $\sim10$ half light radii to the action of baryonic gas condensation into the halo centre. The distribution and movement of gas within galaxies is itself related to the strength of stellar feedback, and so now we compare our fiducial model with the weak feedback model used in \citet{agertz2020}. This model artificially limits the maximum velocity and temperature of \ac{SNe} ejecta to $1000\,\rm{km}\,\rm{s}^{-1}$ and $10^8\,$K, and should be seen as a useful, rather than physical, model that reduces the strength of \ac{SNe}-driven feedback to assess the sensitivity of our results. \par

We contrast our original baryonic physics scheme with this weak feedback model in two example realisations, Halo1459 and Halo600, shown once again at our lower resolution level. This choice represents a lower-mass dwarf halo for which the halo shape was mostly unaffected by the addition of baryonic physics, and a higher-mass dwarf halo which was more highly affected. As in \S\ref{GM}, these simulations are run at our fiducial resolution. Due to the reduced stellar feedback, both galaxies form roughly an order of magnitude more stellar mass and over a longer period of time. Their final stellar masses are unrealistically high compared to their halo mass, and we stress that these models are intended for investigative purposes rather than as a viable alternative physics scheme. Whilst the stellar feedback from these stars are limited, it is still capable of driving temporal fluctuations in the gas density within the inner kpc (of around an order of magnitude). The galaxies also maintain a higher average gas density out to beyond their half light radii ($\sim10\,$kpc, but then converging with the original physics model towards the virial radius). We compare the affects on the final \ac{DM} halo shapes in Figure \ref{fig:fblim_shape}, and on the velocity ansiotropy profiles in Figure \ref{fig:fblim_ansitropy}. \par

In more massive galaxies, a reduction in the stellar feedback leads to a raised central gas density, which in turn acts to gravitationally scatter \ac{DM} particle orbits into rounder and more oblate configurations \citep{2016MNRAS.458.4477T, 2022MNRAS.515.2681C}. The same result is seen in Figure \ref{fig:fblim_shape}, with the axial ratios of the \ac{DM} halo now rounder ($\Delta(x/a)\simeq0.2$) than in the original simulations. The two physics models begin to converge at $a\sim10\,$kpc and beyond. \par

The velocity anisotropy profiles shown in Figure \ref{fig:fblim_ansitropy} reveal that the \ac{DM} anisotropy is almost entirely equivalent between both feedback models, but the stellar profiles are depressed at larger orbital radii in the weak feedback model (from close to the half light radius and outwards). \par

It is important to note that the axial ratio changes caused by this weak feedback model are of a comparable magnitude to the changes caused by a different assembly history in \S\ref{GM} ($\Delta(x/a)\simeq0.2$). Whilst it is true that this weak feedback model is not necessarily realistic, it nevertheless highlights the importance of accounting for the strength of stellar feedback. The combined effects of feedback physics and assembly history on the halo shape could prove difficult to disentangle in practice. \par

\section{Discussion} \label{discussion}

\subsection{A minimum scale for shape transformation}

Our results support previous literature in that pure \ac{DM} halo shapes are prolate in their centres, and become increasingly oblate and round towards their outer edges \citep{2007MNRAS.376..215B, 2015MNRAS.450.2327Z, 2016MNRAS.458.4477T, 2018AAS...23135606W}. As reported elsewhere, this is caused by anisotropic accretion in the early universe followed by more isotropic accretion at late times (see \S\ref{a_anisotropy}). When `baryonic physics' -- gas cooling, star formation and stellar feedback -- are included, we find that the condensation of gas into the halo centre acts to transform the inner prolate shape into a rounder, more oblate shape within $\sim10$ half light radii. In \ac{EDGE}, this effect becomes negligible below certain mass scales ($M_{\rm{200c}} = 1.43\times10^9$\,M$_\odot$; $M_{*} = 3.77\times10^5$\,M$_\odot$). This is due to two reasons: the central gas mass fraction rapidly diminishes with halo mass below this scale, and so does the star formation and subsequently the number of \ac{SNe}. Both of these qualities are required for the gas flows that transform the \ac{DM} halo shape. We predict, therefore, that observations of gas rich ultra-faints should reveal oblate \ac{DM} haloes that are aligned with the inner gas disc (where present), while gas poor ultra-faint dwarf galaxies should reveal primordial triaxial haloes that are prolate on average. We note, however, that our small sample drawn from the \ac{EDGE} simulation suite may not be fully representative of galaxy variety in the field. Other suites which probe similar mass ranges may help to prove this issue further \citep[e.g.][]{2022MNRAS.516.5914C}. \par

A gas disc forms in just one of our simulations (Halo624). For this case, the halo is oblate and aligned with the angular momentum vector of the cool gas, similar to the behaviour reported for more massive disc galaxies \citep[e.g.][]{2008ApJ...681.1076D}. Whilst the \ac{DM} halo shape and rotating stellar component are already established and stable at early times (see \S\ref{shape}), the gas disc only starts to develop at $\sim9$\,Gyr. This indicates that the host halo influences the orientation of the gas disc. Furthermore, it may suggest that the highly oblate halo shape \textit{encourages} the formation of a gas disc at later times due to the torquing of accreted gas (similar to the effect reported in \citealt{2009MNRAS.397...44R}). This will be investigated fully in future work (Rey et al. in prep.). \par

\subsection{A connection between halo shape, velocity anisotropy and assembly}

We find that the triaxiality and velocity anisotropy profiles of a halo are a product of their assembly history, in accordance with prior work on the subject \citep{1997MNRAS.286..865T, 2004MNRAS.354..522M, 2007MNRAS.376.1261M, 2011MNRAS.416.1377V, 2012JCAP...10..049S, 2018MNRAS.476.1796S, 2018MNRAS.481..414G, 2023arXiv230208853C}. The inner halo ($\sim0\mathit{\textnormal{-}}1\,$kpc) is typically constructed from anisotropic accretion at high redshift, and tends towards a triaxial prolate shape. This can then be modified by inward gas condensation. The outer halo ($>1\,$kpc) is highly dependent on the major mergers that the halo experiences at intermediate redshift ($z\sim5\mathit{\textnormal{-}}2$). A halo dominated by major mergers can undergo a flattening in the plane of the merger, yielding an oblate outer halo shape. This is due both to the gravitational elongation provided by the merger, and the merger debris being deposited in an oblate track around the infall plane. This phenomena relies on a high angular momentum of the merger, and our limited sample of simulations is insufficient to investigate results across a spectrum of angular momenta. Conversely, a halo with no major merger will retain its primordial prolate outer shape, and the shape is less likely to be modified by gas flows at these higher radii. \par

The velocity anisotropy of the stars is more radially biased than the corresponding \ac{DM}, except in the presence of rotating stellar features. This is due to the contamination of stars from ex-situ origins, which increasingly dominate at higher radii and favour radial velocity distributions. We discussed the physical reasons for this result in \S\ref{stars_DM}. The same may not be true for dwarf galaxies at higher mass scales, where the stellar outskirts are polluted by disc-heated stars \citep[dwarf spheroidals, e.g.][]{1996ApJ...467L..13M, 2020ApJ...900..163K, 2022ApJ...931..152K}. We will begin to probe this mass regime in future \ac{EDGE} simulations. \par

\subsection{Implications for obtaining the mass distribution in nearby dwarfs}

The dynamical mass distribution of nearby dwarf galaxies is often inferred from the motions of their stars (\citealp{2008ApJ...682..835Z, 2014JCAP...02..023H,2019MNRAS.484.1401R,2020MNRAS.498..144G,2021arXiv211209374Z} and see \citealp{2003MNRAS.343..401L, 2017MNRAS.471.4541R}). In particular, \citet{2017MNRAS.471.4541R} show that, on average, $>10,000$ velocities are needed to detect deviations from spherical symmetry with any degree of confidence. For fewer measurements than this, the biases introduced by a triaxial halo are typically less than the formal uncertainties in fitting the model. The advent of multi-object spectrographs on the 30\,m class telescopes such as the ELT \citep{2021Msngr.182...27M} may eventually provide the velocity measurements needed to probe halo triaxiality in ultra-faints with $M_{*}\sim10^5\,\rm{M}_{\odot}$, at which point cosmologically-motivated priors would be beneficial. It may never be possible to perform such measurements in ultra-faints with $M_{*}\sim10^3\,\rm{M}_{\odot}$. \par

However, interesting constraints on the spherically averaged radial density profile can be obtained with even of $\mathcal{O}(100)$ tracer stars \citep[e.g.][]{2021MNRAS.505.5686C,2021arXiv211209374Z}. Such constraints require us to measure or marginalise over the velocity anisotropy profile of the stars -- a significant source of uncertainty \citep[e.g.][]{2017MNRAS.471.4541R}. Our results here suggest that such models could employ a strong prior that the velocity anisotropy is isotropic in the centre, becoming highly radial ($\tilde{\beta} > 0.5$) beyond $r \gtrsim 3 R_{\rm half}$. We will explore in future work how this prior impacts our ability to measure the mass distribution in the smallest galaxies.\par

While we are unlikely to have enough stellar radial velocities to directly probe halo shapes in the smallest galaxies, combining stellar and gas kinematics for more massive nearby dwarfs shows promise. \citet{2021MNRAS.500..410L} investigated the central density and shape of the isolated dwarf irregular galaxy WLM using joint stellar and gas kinematics. WLM, with a stellar mass of $M_* \approx 1\mathit{\textnormal{-}}4\times10^7\,\text{M}_{\odot}$ \citep{2007ApJ...656..818J, 2012AJ....144....4M} and a halo mass of order $M_{\rm{200c}} \sim 10^{10}$\,M$_\odot$ \citep{2012ApJ...750...33L,2017MNRAS.467.2019R}, is in a higher mass category than the \ac{EDGE} simulations presented here. It also has a correspondingly richer star formation history \citep[i.e.][]{2019MNRAS.490.5538A}. \citet{2017MNRAS.467.2019R} and \citet{2021MNRAS.500..410L} find evidence favouring a \ac{DM} core, consistent with having formed through bursty stellar feedback. \citet{2021MNRAS.500..410L} also find evidence for a prolate halo shape. This is particularly interesting since WLM has a clear gas disc and so one might expect its inner halo shape to be oblate and aligned with the disc, as in Halo624. It remains to be seen if this prolate shape is a challenge for our current galaxy formation models in a \ac{LCDM} cosmology. We will address this in future work as we simulate a larger sample of more massive dwarfs in \ac{EDGE}.\par

\section{Conclusions} \label{conclusion}

We have analysed five ultra-faint dwarf galaxies over the mass range $1.3\times10^9 \leqslant M_{\rm 200c}/\text{M}_{\odot} \leqslant 3.2\times10^9$, run with a minimum \ac{DM} mass resolution of $120\,\text{M}_{\odot}$ as part of the \ac{EDGE} project. This resolution is sufficient to analyse the shape of the \ac{DM} profile at radii $< 100$\,pc, which is within the half light radii of the dwarfs in this mass regime. \par

For dwarfs of mass $M_{\rm 200c} > 3\times10^9$, the condensation of baryons into the centres of \ac{DM} haloes transforms an otherwise prolate shape into a more triaxial or oblate shape -- similar to the effect reported in more massive galaxies. This transformation occurs on scales $\sim10$ greater than the 3D stellar half light radius. Large amounts of inflowing gas can contribute to transforming the halo shape -- regardless of whether there is any associated \ac{DM} heating in the density profile. However, \ac{DM} heating may make the central halo more vulnerable to the impact of major mergers and gas ejecta, which can themselves modify the halo shape. Further investigation with more idealised setups is needed to confirm the magnitude of these effects. \par

Significant transformations in the \ac{DM} halo shape occur even in dwarf galaxies with present-day gas-to-\ac{DM} mass fractions as low as $\sim7\times10^{-4}$. This is because the central gas fractions were many times greater in the early universe, and the shape transformations occur predominantly at this primordial epoch. \par

For dwarfs of mass $M_{\rm 200c} \leqslant 1.5\times10^9$, the addition of baryonic physics yields no appreciable difference in the \ac{DM} halo shapes. This because there is insufficient condensation of barons into the halo centres. We predict that future observations of gas-poor ultra-faint dwarfs will find an increased prevalence of primordial, prolate \ac{DM} haloes. \par

Analysis of the velocity anisotropy reveals similar trends to those reported elsewhere in the literature, with isotropic halo centres that are increasingly radially anisotropic at higher radii. However, the anisotropy profile slope of the stellar component is far greater than that of the \ac{DM} in most cases. This may be important for designing the priors used in mass modelling of the smallest galaxies. \par

\section{Acknowledgements}

We thank the anonymous referee for their helpful and insightful comments.

MO acknowledges the UKRI Science and Technology Facilities Council (STFC) for support (grant ST/R505134/1), and funding from the European Research Council (ERC) under the European Union’s Horizon 2020 research and innovation programme (grant agreement No. 852839). ET acknowledges
the UKRI Science and Technology Facilities Council (STFC) for support (grant ST/V50712X/1). MR is supported by the Beecroft Fellowship funded by Adrian Beecroft. OA acknowledges financial support from the Knut and Alice Wallenberg Foundation and the Swedish Research Council (grant 2019-04659).
This work was performed using the DiRAC Data Intensive service at Leicester, operated by the University of Leicester IT Services, which forms part of the STFC DiRAC HPC Facility (www.dirac.ac.uk). The equipment was funded by BEIS capital funding via STFC capital grants ST/K000373/1 and ST/R002363/1 and STFC DiRAC Operations grant ST/R001014/1. DiRAC is part of the National e-Infrastructure.
The authors acknowledge the use of the Surrey Eureka supercomputer facility and associated support
services.

\section*{Data availability}
Data is available upon reasonable request.



\bibliographystyle{mnras}
\bibliography{main} 

\begin{thebibliography}{}
\makeatletter
\relax
\def\mn@urlcharsother{\let\do\@makeother \do\$\do\&\do\#\do\^\do\_\do\%\do\~}
\def\mn@doi{\begingroup\mn@urlcharsother \@ifnextchar [ {\mn@doi@}
  {\mn@doi@[]}}
\def\mn@doi@[#1]#2{\def\@tempa{#1}\ifx\@tempa\@empty \href
  {http://dx.doi.org/#2} {doi:#2}\else \href {http://dx.doi.org/#2} {#1}\fi
  \endgroup}
\def\mn@eprint#1#2{\mn@eprint@#1:#2::\@nil}
\def\mn@eprint@arXiv#1{\href {http://arxiv.org/abs/#1} {{\tt arXiv:#1}}}
\def\mn@eprint@dblp#1{\href {http://dblp.uni-trier.de/rec/bibtex/#1.xml}
  {dblp:#1}}
\def\mn@eprint@#1:#2:#3:#4\@nil{\def\@tempa {#1}\def\@tempb {#2}\def\@tempc
  {#3}\ifx \@tempc \@empty \let \@tempc \@tempb \let \@tempb \@tempa \fi \ifx
  \@tempb \@empty \def\@tempb {arXiv}\fi \@ifundefined
  {mn@eprint@\@tempb}{\@tempb:\@tempc}{\expandafter \expandafter \csname
  mn@eprint@\@tempb\endcsname \expandafter{\@tempc}}}

\bibitem[\protect\citeauthoryear{{Agertz} et~al.,}{{Agertz}
  et~al.}{2020}]{agertz2020}
{Agertz} O.,  et~al., 2020, \mn@doi [\mnras] {10.1093/mnras/stz3053}, \href
  {https://ui.adsabs.harvard.edu/abs/2020MNRAS.491.1656A} {491, 1656}

\bibitem[\protect\citeauthoryear{{Albers} et~al.,}{{Albers}
  et~al.}{2019}]{2019MNRAS.490.5538A}
{Albers} S.~M.,  et~al., 2019, \mn@doi [\mnras] {10.1093/mnras/stz2903}, \href
  {https://ui.adsabs.harvard.edu/abs/2019MNRAS.490.5538A} {490, 5538}

\bibitem[\protect\citeauthoryear{{Allgood}, {Flores}, {Primack}, {Kravtsov},
  {Wechsler}, {Faltenbacher}  \& {Bullock}}{{Allgood}
  et~al.}{2006}]{2006MNRAS.367.1781A}
{Allgood} B.,  {Flores} R.~A.,  {Primack} J.~R.,  {Kravtsov} A.~V.,  {Wechsler}
  R.~H.,  {Faltenbacher} A.,   {Bullock} J.~S.,  2006, \mn@doi [\mnras]
  {10.1111/j.1365-2966.2006.10094.x}, \href
  {http://adsabs.harvard.edu/abs/2006MNRAS.367.1781A} {367, 1781}

\bibitem[\protect\citeauthoryear{{Bardeen}, {Bond}, {Kaiser}  \&
  {Szalay}}{{Bardeen} et~al.}{1986}]{1986ApJ...304...15B}
{Bardeen} J.~M.,  {Bond} J.~R.,  {Kaiser} N.,   {Szalay} A.~S.,  1986, \mn@doi
  [\apj] {10.1086/164143}, \href
  {https://ui.adsabs.harvard.edu/abs/1986ApJ...304...15B} {304, 15}

\bibitem[\protect\citeauthoryear{{Bett}, {Eke}, {Frenk}, {Jenkins}, {Helly}  \&
  {Navarro}}{{Bett} et~al.}{2007}]{2007MNRAS.376..215B}
{Bett} P.,  {Eke} V.,  {Frenk} C.~S.,  {Jenkins} A.,  {Helly} J.,   {Navarro}
  J.,  2007, \mn@doi [\mnras] {10.1111/j.1365-2966.2007.11432.x}, \href
  {https://ui.adsabs.harvard.edu/abs/2007MNRAS.376..215B} {376, 215}

\bibitem[\protect\citeauthoryear{{Binney}}{{Binney}}{1980}]{1980MNRAS.190..873B}
{Binney} J.,  1980, \mn@doi [\mnras] {10.1093/mnras/190.4.873}, \href
  {https://ui.adsabs.harvard.edu/abs/1980MNRAS.190..873B} {190, 873}

\bibitem[\protect\citeauthoryear{{Binney} \& {Tremaine}}{{Binney} \&
  {Tremaine}}{2008}]{2008gady.book.....B}
{Binney} J.,  {Tremaine} S.,  2008, {Galactic Dynamics: Second Edition}

\bibitem[\protect\citeauthoryear{{Brinckmann}, {Zavala}, {Rapetti}, {Hansen}
  \& {Vogelsberger}}{{Brinckmann} et~al.}{2018}]{2018MNRAS.474..746B}
{Brinckmann} T.,  {Zavala} J.,  {Rapetti} D.,  {Hansen} S.~H.,   {Vogelsberger}
  M.,  2018, \mn@doi [\mnras] {10.1093/mnras/stx2782}, \href
  {https://ui.adsabs.harvard.edu/abs/2018MNRAS.474..746B} {474, 746}

\bibitem[\protect\citeauthoryear{{Bruderer}, {Read}, {Coles}, {Leier}, {Falco},
  {Ferreras}  \& {Saha}}{{Bruderer} et~al.}{2016}]{2016MNRAS.456..870B}
{Bruderer} C.,  {Read} J.~I.,  {Coles} J.~P.,  {Leier} D.,  {Falco} E.~E.,
  {Ferreras} I.,   {Saha} P.,  2016, \mn@doi [\mnras] {10.1093/mnras/stv2582},
  \href {https://ui.adsabs.harvard.edu/abs/2016MNRAS.456..870B} {456, 870}

\bibitem[\protect\citeauthoryear{{Bryan}, {Kay}, {Duffy}, {Schaye}, {Dalla
  Vecchia}  \& {Booth}}{{Bryan} et~al.}{2013}]{2013MNRAS.429.3316B}
{Bryan} S.~E.,  {Kay} S.~T.,  {Duffy} A.~R.,  {Schaye} J.,  {Dalla Vecchia} C.,
    {Booth} C.~M.,  2013, \mn@doi [\mnras] {10.1093/mnras/sts587}, \href
  {https://ui.adsabs.harvard.edu/abs/2013MNRAS.429.3316B} {429, 3316}

\bibitem[\protect\citeauthoryear{{Calura} et~al.,}{{Calura}
  et~al.}{2022}]{2022MNRAS.516.5914C}
{Calura} F.,  et~al., 2022, \mn@doi [\mnras] {10.1093/mnras/stac2387}, \href
  {https://ui.adsabs.harvard.edu/abs/2022MNRAS.516.5914C} {516, 5914}

\bibitem[\protect\citeauthoryear{{Cataldi}, {Pedrosa}, {Tissera}  \&
  {Artale}}{{Cataldi} et~al.}{2021}]{2021MNRAS.501.5679C}
{Cataldi} P.,  {Pedrosa} S.~E.,  {Tissera} P.~B.,   {Artale} M.~C.,  2021,
  \mn@doi [\mnras] {10.1093/mnras/staa3988}, \href
  {https://ui.adsabs.harvard.edu/abs/2021MNRAS.501.5679C} {501, 5679}

\bibitem[\protect\citeauthoryear{{Cataldi} et~al.,}{{Cataldi}
  et~al.}{2023}]{2023arXiv230208853C}
{Cataldi} P.,  et~al., 2023, \mn@doi [\mnras] {10.1093/mnras/stad1601}, \href
  {https://ui.adsabs.harvard.edu/abs/2023MNRAS.523.1919C} {523, 1919}

\bibitem[\protect\citeauthoryear{{Ceverino}, {Primack}  \& {Dekel}}{{Ceverino}
  et~al.}{2015}]{2015MNRAS.453..408C}
{Ceverino} D.,  {Primack} J.,   {Dekel} A.,  2015, \mn@doi [\mnras]
  {10.1093/mnras/stv1603}, \href
  {https://ui.adsabs.harvard.edu/abs/2015MNRAS.453..408C} {453, 408}

\bibitem[\protect\citeauthoryear{{Chua}, {Pillepich}, {Vogelsberger}  \&
  {Hernquist}}{{Chua} et~al.}{2019}]{2019MNRAS.484..476C}
{Chua} K. T.~E.,  {Pillepich} A.,  {Vogelsberger} M.,   {Hernquist} L.,  2019,
  \mn@doi [\mnras] {10.1093/mnras/sty3531}, \href
  {https://ui.adsabs.harvard.edu/abs/2019MNRAS.484..476C} {484, 476}

\bibitem[\protect\citeauthoryear{{Chua}, {Vogelsberger}, {Pillepich}  \&
  {Hernquist}}{{Chua} et~al.}{2022}]{2022MNRAS.515.2681C}
{Chua} K. T.~E.,  {Vogelsberger} M.,  {Pillepich} A.,   {Hernquist} L.,  2022,
  \mn@doi [\mnras] {10.1093/mnras/stac1897}, \href
  {https://ui.adsabs.harvard.edu/abs/2022MNRAS.515.2681C} {515, 2681}

\bibitem[\protect\citeauthoryear{{Collins} et~al.,}{{Collins}
  et~al.}{2021}]{2021MNRAS.505.5686C}
{Collins} M. L.~M.,  et~al., 2021, \mn@doi [\mnras] {10.1093/mnras/stab1624},
  \href {https://ui.adsabs.harvard.edu/abs/2021MNRAS.505.5686C} {505, 5686}

\bibitem[\protect\citeauthoryear{{Cooper} et~al.,}{{Cooper}
  et~al.}{2010}]{2010MNRAS.406..744C}
{Cooper} A.~P.,  et~al., 2010, \mn@doi [\mnras]
  {10.1111/j.1365-2966.2010.16740.x}, \href
  {https://ui.adsabs.harvard.edu/abs/2010MNRAS.406..744C} {406, 744}

\bibitem[\protect\citeauthoryear{{Cooper}, {Parry}, {Lowing}, {Cole}  \&
  {Frenk}}{{Cooper} et~al.}{2015}]{2015MNRAS.454.3185C}
{Cooper} A.~P.,  {Parry} O.~H.,  {Lowing} B.,  {Cole} S.,   {Frenk} C.,  2015,
  \mn@doi [\mnras] {10.1093/mnras/stv2057}, \href
  {https://ui.adsabs.harvard.edu/abs/2015MNRAS.454.3185C} {454, 3185}

\bibitem[\protect\citeauthoryear{{Dav{\'e}}, {Spergel}, {Steinhardt}  \&
  {Wandelt}}{{Dav{\'e}} et~al.}{2001}]{2001ApJ...547..574D}
{Dav{\'e}} R.,  {Spergel} D.~N.,  {Steinhardt} P.~J.,   {Wandelt} B.~D.,  2001,
  \mn@doi [\apj] {10.1086/318417}, \href
  {https://ui.adsabs.harvard.edu/abs/2001ApJ...547..574D} {547, 574}

\bibitem[\protect\citeauthoryear{{Debattista}, {Moore}, {Quinn}, {Kazantzidis},
  {Maas}, {Mayer}, {Read}  \& {Stadel}}{{Debattista}
  et~al.}{2008}]{2008ApJ...681.1076D}
{Debattista} V.~P.,  {Moore} B.,  {Quinn} T.,  {Kazantzidis} S.,  {Maas} R.,
  {Mayer} L.,  {Read} J.,   {Stadel} J.,  2008, \mn@doi [\apj]
  {10.1086/587977}, \href
  {https://ui.adsabs.harvard.edu/abs/2008ApJ...681.1076D} {681, 1076}

\bibitem[\protect\citeauthoryear{{Deg} \& {Widrow}}{{Deg} \&
  {Widrow}}{2013}]{2013MNRAS.428..912D}
{Deg} N.,  {Widrow} L.,  2013, \mn@doi [\mnras] {10.1093/mnras/sts089}, \href
  {https://ui.adsabs.harvard.edu/abs/2013MNRAS.428..912D} {428, 912}

\bibitem[\protect\citeauthoryear{{Despali}, {Walls}, {Vegetti}, {Sparre},
  {Vogelsberger}  \& {Zavala}}{{Despali} et~al.}{2022}]{2022arXiv220412502D}
{Despali} G.,  {Walls} L.~G.,  {Vegetti} S.,  {Sparre} M.,  {Vogelsberger} M.,
   {Zavala} J.,  2022, \mn@doi [\mnras] {10.1093/mnras/stac2521}, \href
  {https://ui.adsabs.harvard.edu/abs/2022MNRAS.516.4543D} {516, 4543}

\bibitem[\protect\citeauthoryear{{Dubinski}}{{Dubinski}}{1994}]{1994ApJ...431..617D}
{Dubinski} J.,  1994, \mn@doi [\apj] {10.1086/174512}, \href
  {https://ui.adsabs.harvard.edu/abs/1994ApJ...431..617D} {431, 617}

\bibitem[\protect\citeauthoryear{{Dubinski} \& {Carlberg}}{{Dubinski} \&
  {Carlberg}}{1991}]{1991ApJ...378..496D}
{Dubinski} J.,  {Carlberg} R.~G.,  1991, \mn@doi [\apj] {10.1086/170451}, \href
  {http://adsabs.harvard.edu/abs/1991ApJ...378..496D} {378, 496}

\bibitem[\protect\citeauthoryear{{Dutta Chowdhury}, {van den Bosch}, {van
  Dokkum}, {Robles}, {Schive}  \& {Chiueh}}{{Dutta Chowdhury}
  et~al.}{2023}]{2023ApJ...949...68D}
{Dutta Chowdhury} D.,  {van den Bosch} F.~C.,  {van Dokkum} P.,  {Robles}
  V.~H.,  {Schive} H.-Y.,   {Chiueh} T.,  2023, \mn@doi [\apj]
  {10.3847/1538-4357/acc73d}, \href
  {https://ui.adsabs.harvard.edu/abs/2023ApJ...949...68D} {949, 68}

\bibitem[\protect\citeauthoryear{{Eisenstein} \& {Hut}}{{Eisenstein} \&
  {Hut}}{1998}]{hop}
{Eisenstein} D.~J.,  {Hut} P.,  1998, \mn@doi [\apj] {10.1086/305535}, \href
  {https://ui.adsabs.harvard.edu/abs/1998ApJ...498..137E} {498, 137}

\bibitem[\protect\citeauthoryear{{Fischer} \& {Valenzuela}}{{Fischer} \&
  {Valenzuela}}{2023}]{2022arXiv220911244F}
{Fischer} M.~S.,  {Valenzuela} L.~M.,  2023, \mn@doi [\aap]
  {10.1051/0004-6361/202245031}, \href
  {https://ui.adsabs.harvard.edu/abs/2023A&A...670A.120F} {670, A120}

\bibitem[\protect\citeauthoryear{{Franx}, {Illingworth}  \& {de Zeeuw}}{{Franx}
  et~al.}{1991}]{1991ApJ...383..112F}
{Franx} M.,  {Illingworth} G.,   {de Zeeuw} T.,  1991, \mn@doi [\apj]
  {10.1086/170769}, \href
  {https://ui.adsabs.harvard.edu/abs/1991ApJ...383..112F} {383, 112}

\bibitem[\protect\citeauthoryear{{Frenk}, {White}, {Davis}  \&
  {Efstathiou}}{{Frenk} et~al.}{1988}]{1988ApJ...327..507F}
{Frenk} C.~S.,  {White} S. D.~M.,  {Davis} M.,   {Efstathiou} G.,  1988,
  \mn@doi [\apj] {10.1086/166213}, \href
  {https://ui.adsabs.harvard.edu/abs/1988ApJ...327..507F} {327, 507}

\bibitem[\protect\citeauthoryear{{Ganeshaiah Veena}, {Cautun}, {van de
  Weygaert}, {Tempel}, {Jones}, {Rieder}  \& {Frenk}}{{Ganeshaiah Veena}
  et~al.}{2018}]{2018MNRAS.481..414G}
{Ganeshaiah Veena} P.,  {Cautun} M.,  {van de Weygaert} R.,  {Tempel} E.,
  {Jones} B. J.~T.,  {Rieder} S.,   {Frenk} C.~S.,  2018, \mn@doi [\mnras]
  {10.1093/mnras/sty2270}, \href
  {https://ui.adsabs.harvard.edu/abs/2018MNRAS.481..414G} {481, 414}

\bibitem[\protect\citeauthoryear{{Genina} et~al.,}{{Genina}
  et~al.}{2020}]{2020MNRAS.498..144G}
{Genina} A.,  et~al., 2020, \mn@doi [\mnras] {10.1093/mnras/staa2352}, \href
  {https://ui.adsabs.harvard.edu/abs/2020MNRAS.498..144G} {498, 144}

\bibitem[\protect\citeauthoryear{{Gerhard} \& {Binney}}{{Gerhard} \&
  {Binney}}{1985}]{1985MNRAS.216..467G}
{Gerhard} O.~E.,  {Binney} J.,  1985, \mn@doi [\mnras]
  {10.1093/mnras/216.2.467}, \href
  {https://ui.adsabs.harvard.edu/abs/1985MNRAS.216..467G} {216, 467}

\bibitem[\protect\citeauthoryear{{Goater} et~al.,}{{Goater}
  et~al.}{2023}]{2023arXiv230705130G}
{Goater} A.,  et~al., 2023, \mn@doi [arXiv e-prints]
  {10.48550/arXiv.2307.05130}, \href
  {https://ui.adsabs.harvard.edu/abs/2023arXiv230705130G} {p. arXiv:2307.05130}

\bibitem[\protect\citeauthoryear{{Gustafsson}, {Fairbairn}  \&
  {Sommer-Larsen}}{{Gustafsson} et~al.}{2006}]{2006PhRvD..74l3522G}
{Gustafsson} M.,  {Fairbairn} M.,   {Sommer-Larsen} J.,  2006, \mn@doi [\prd]
  {10.1103/PhysRevD.74.123522}, \href
  {http://adsabs.harvard.edu/abs/2006PhRvD..74l3522G} {74, 123522}

\bibitem[\protect\citeauthoryear{{Helmi}}{{Helmi}}{2004}]{2004ApJ...610L..97H}
{Helmi} A.,  2004, \mn@doi [\apjl] {10.1086/423340}, \href
  {https://ui.adsabs.harvard.edu/abs/2004ApJ...610L..97H} {610, L97}

\bibitem[\protect\citeauthoryear{{Hernquist}}{{Hernquist}}{1990}]{1990ApJ...356..359H}
{Hernquist} L.,  1990, \mn@doi [\apj] {10.1086/168845}, \href
  {http://adsabs.harvard.edu/abs/1990ApJ...356..359H} {356, 359}

\bibitem[\protect\citeauthoryear{{Hunter}}{{Hunter}}{2014}]{2014JCAP...02..023H}
{Hunter} D.~R.,  2014, \mn@doi [\jcap] {10.1088/1475-7516/2014/02/023}, \href
  {https://ui.adsabs.harvard.edu/abs/2014JCAP...02..023H} {2014, 023}

\bibitem[\protect\citeauthoryear{{Ibata}, {Lewis}, {Irwin}, {Totten}  \&
  {Quinn}}{{Ibata} et~al.}{2001}]{2001ApJ...551..294I}
{Ibata} R.,  {Lewis} G.~F.,  {Irwin} M.,  {Totten} E.,   {Quinn} T.,  2001,
  \mn@doi [\apj] {10.1086/320060}, \href
  {https://ui.adsabs.harvard.edu/abs/2001ApJ...551..294I} {551, 294}

\bibitem[\protect\citeauthoryear{{Jackson}, {Skillman}, {Gehrz}, {Polomski}  \&
  {Woodward}}{{Jackson} et~al.}{2007}]{2007ApJ...656..818J}
{Jackson} D.~C.,  {Skillman} E.~D.,  {Gehrz} R.~D.,  {Polomski} E.,
  {Woodward} C.~E.,  2007, \mn@doi [\apj] {10.1086/510354}, \href
  {https://ui.adsabs.harvard.edu/abs/2007ApJ...656..818J} {656, 818}

\bibitem[\protect\citeauthoryear{{Jing} \& {Suto}}{{Jing} \&
  {Suto}}{2002}]{2002ApJ...574..538J}
{Jing} Y.~P.,  {Suto} Y.,  2002, \mn@doi [\apj] {10.1086/341065}, \href
  {http://adsabs.harvard.edu/abs/2002ApJ...574..538J} {574, 538}

\bibitem[\protect\citeauthoryear{{Kado-Fong}, {Greene}, {Huang}, {Beaton},
  {Goulding}  \& {Komiyama}}{{Kado-Fong} et~al.}{2020}]{2020ApJ...900..163K}
{Kado-Fong} E.,  {Greene} J.~E.,  {Huang} S.,  {Beaton} R.,  {Goulding} A.~D.,
   {Komiyama} Y.,  2020, \mn@doi [\apj] {10.3847/1538-4357/abacc2}, \href
  {https://ui.adsabs.harvard.edu/abs/2020ApJ...900..163K} {900, 163}

\bibitem[\protect\citeauthoryear{{Kado-Fong} et~al.,}{{Kado-Fong}
  et~al.}{2022}]{2022ApJ...931..152K}
{Kado-Fong} E.,  et~al., 2022, \mn@doi [\apj] {10.3847/1538-4357/ac6c88}, \href
  {https://ui.adsabs.harvard.edu/abs/2022ApJ...931..152K} {931, 152}

\bibitem[\protect\citeauthoryear{{Kasun} \& {Evrard}}{{Kasun} \&
  {Evrard}}{2005}]{2005ApJ...629..781K}
{Kasun} S.~F.,  {Evrard} A.~E.,  2005, \mn@doi [\apj]
  {10.1086/43081110.48550/arXiv.astro-ph/0408056}, \href
  {https://ui.adsabs.harvard.edu/abs/2005ApJ...629..781K} {629, 781}

\bibitem[\protect\citeauthoryear{{Katz}}{{Katz}}{1991}]{1991ApJ...368..325K}
{Katz} N.,  1991, \mn@doi [\apj] {10.1086/169696}, \href
  {http://adsabs.harvard.edu/abs/1991ApJ...368..325K} {368, 325}

\bibitem[\protect\citeauthoryear{{Katz} \& {Gunn}}{{Katz} \&
  {Gunn}}{1991}]{1991ApJ...377..365K}
{Katz} N.,  {Gunn} J.~E.,  1991, \mn@doi [\apj] {10.1086/170367}, \href
  {https://ui.adsabs.harvard.edu/abs/1991ApJ...377..365K} {377, 365}

\bibitem[\protect\citeauthoryear{{Kazantzidis}, {Abadi}  \&
  {Navarro}}{{Kazantzidis} et~al.}{2010}]{2010ApJ...720L..62K}
{Kazantzidis} S.,  {Abadi} M.~G.,   {Navarro} J.~F.,  2010, \mn@doi [\apjl]
  {10.1088/2041-8205/720/1/L62}, \href
  {https://ui.adsabs.harvard.edu/abs/2010ApJ...720L..62K} {720, L62}

\bibitem[\protect\citeauthoryear{{Kirby}, {Bullock}, {Boylan-Kolchin},
  {Kaplinghat}  \& {Cohen}}{{Kirby} et~al.}{2014}]{2014MNRAS.439.1015K}
{Kirby} E.~N.,  {Bullock} J.~S.,  {Boylan-Kolchin} M.,  {Kaplinghat} M.,
  {Cohen} J.~G.,  2014, \mn@doi [\mnras] {10.1093/mnras/stu025}, \href
  {https://ui.adsabs.harvard.edu/abs/2014MNRAS.439.1015K} {439, 1015}

\bibitem[\protect\citeauthoryear{{Knollmann} \& {Knebe}}{{Knollmann} \&
  {Knebe}}{2009}]{2009ApJS..182..608K}
{Knollmann} S.~R.,  {Knebe} A.,  2009, \mn@doi [\apjs]
  {10.1088/0067-0049/182/2/608}, \href
  {https://ui.adsabs.harvard.edu/abs/2009ApJS..182..608K} {182, 608}

\bibitem[\protect\citeauthoryear{{Leaman} et~al.,}{{Leaman}
  et~al.}{2012}]{2012ApJ...750...33L}
{Leaman} R.,  et~al., 2012, \mn@doi [\apj] {10.1088/0004-637X/750/1/33}, \href
  {https://ui.adsabs.harvard.edu/abs/2012ApJ...750...33L} {750, 33}

\bibitem[\protect\citeauthoryear{{Lemze} et~al.,}{{Lemze}
  et~al.}{2012}]{2012ApJ...752..141L}
{Lemze} D.,  et~al., 2012, \mn@doi [\apj] {10.1088/0004-637X/752/2/141}, \href
  {https://ui.adsabs.harvard.edu/abs/2012ApJ...752..141L} {752, 141}

\bibitem[\protect\citeauthoryear{{Leung}, {Leaman}, {Battaglia}, {van de Ven},
  {Brooks}, {Pe{\~n}arrubia}  \& {Venn}}{{Leung}
  et~al.}{2021}]{2021MNRAS.500..410L}
{Leung} G. Y.~C.,  {Leaman} R.,  {Battaglia} G.,  {van de Ven} G.,  {Brooks}
  A.~M.,  {Pe{\~n}arrubia} J.,   {Venn} K.~A.,  2021, \mn@doi [\mnras]
  {10.1093/mnras/staa3107}, \href
  {https://ui.adsabs.harvard.edu/abs/2021MNRAS.500..410L} {500, 410}

\bibitem[\protect\citeauthoryear{{{\L}okas} \& {Mamon}}{{{\L}okas} \&
  {Mamon}}{2003}]{2003MNRAS.343..401L}
{{\L}okas} E.~L.,  {Mamon} G.~A.,  2003, \mn@doi [\mnras]
  {10.1046/j.1365-8711.2003.06684.x}, \href
  {https://ui.adsabs.harvard.edu/abs/2003MNRAS.343..401L} {343, 401}

\bibitem[\protect\citeauthoryear{{Ludlow}, {Navarro}, {White},
  {Boylan-Kolchin}, {Springel}, {Jenkins}  \& {Frenk}}{{Ludlow}
  et~al.}{2011}]{2011MNRAS.415.3895L}
{Ludlow} A.~D.,  {Navarro} J.~F.,  {White} S. D.~M.,  {Boylan-Kolchin} M.,
  {Springel} V.,  {Jenkins} A.,   {Frenk} C.~S.,  2011, \mn@doi [\mnras]
  {10.1111/j.1365-2966.2011.19008.x}, \href
  {https://ui.adsabs.harvard.edu/abs/2011MNRAS.415.3895L} {415, 3895}

\bibitem[\protect\citeauthoryear{{Lux}, {Read}, {Lake}  \& {Johnston}}{{Lux}
  et~al.}{2012}]{2012MNRAS.424L..16L}
{Lux} H.,  {Read} J.~I.,  {Lake} G.,   {Johnston} K.~V.,  2012, \mn@doi
  [\mnras] {10.1111/j.1745-3933.2012.01276.x}, \href
  {https://ui.adsabs.harvard.edu/abs/2012MNRAS.424L..16L} {424, L16}

\bibitem[\protect\citeauthoryear{{Macci{\`o}}, {Dutton}  \& {van den
  Bosch}}{{Macci{\`o}} et~al.}{2008}]{2008MNRAS.391.1940M}
{Macci{\`o}} A.~V.,  {Dutton} A.~A.,   {van den Bosch} F.~C.,  2008, \mn@doi
  [\mnras] {10.1111/j.1365-2966.2008.14029.x}, \href
  {https://ui.adsabs.harvard.edu/abs/2008MNRAS.391.1940M} {391, 1940}

\bibitem[\protect\citeauthoryear{{Marconi} et~al.,}{{Marconi}
  et~al.}{2021}]{2021Msngr.182...27M}
{Marconi} A.,  et~al., 2021, \mn@doi [The Messenger] {10.18727/0722-6691/5219},
  \href {https://ui.adsabs.harvard.edu/abs/2021Msngr.182...27M} {182, 27}

\bibitem[\protect\citeauthoryear{{McConnachie}}{{McConnachie}}{2012}]{2012AJ....144....4M}
{McConnachie} A.~W.,  2012, \mn@doi [\aj] {10.1088/0004-6256/144/1/4}, \href
  {https://ui.adsabs.harvard.edu/abs/2012AJ....144....4M} {144, 4}

\bibitem[\protect\citeauthoryear{{McMillan}, {Athanassoula}  \&
  {Dehnen}}{{McMillan} et~al.}{2007}]{2007MNRAS.376.1261M}
{McMillan} P.~J.,  {Athanassoula} E.,   {Dehnen} W.,  2007, \mn@doi [\mnras]
  {10.1111/j.1365-2966.2007.11516.x}, \href
  {https://ui.adsabs.harvard.edu/abs/2007MNRAS.376.1261M} {376, 1261}

\bibitem[\protect\citeauthoryear{{Merritt} \& {Valluri}}{{Merritt} \&
  {Valluri}}{1999}]{1999AJ....118.1177M}
{Merritt} D.,  {Valluri} M.,  1999, \mn@doi [\aj] {10.1086/301012}, \href
  {https://ui.adsabs.harvard.edu/abs/1999AJ....118.1177M} {118, 1177}

\bibitem[\protect\citeauthoryear{{Minniti} \& {Zijlstra}}{{Minniti} \&
  {Zijlstra}}{1996}]{1996ApJ...467L..13M}
{Minniti} D.,  {Zijlstra} A.~A.,  1996, \mn@doi [\apjl] {10.1086/310189}, \href
  {https://ui.adsabs.harvard.edu/abs/1996ApJ...467L..13M} {467, L13}

\bibitem[\protect\citeauthoryear{{Moore}, {Kazantzidis}, {Diemand}  \&
  {Stadel}}{{Moore} et~al.}{2004}]{2004MNRAS.354..522M}
{Moore} B.,  {Kazantzidis} S.,  {Diemand} J.,   {Stadel} J.,  2004, \mn@doi
  [\mnras] {10.1111/j.1365-2966.2004.08211.x}, \href
  {https://ui.adsabs.harvard.edu/abs/2004MNRAS.354..522M} {354, 522}

\bibitem[\protect\citeauthoryear{{Navarro}, {Frenk}  \& {White}}{{Navarro}
  et~al.}{1996}]{1996ApJ...462..563N}
{Navarro} J.~F.,  {Frenk} C.~S.,   {White} S. D.~M.,  1996, \mn@doi [\apj]
  {10.1086/177173}, \href
  {https://ui.adsabs.harvard.edu/abs/1996ApJ...462..563N} {462, 563}

\bibitem[\protect\citeauthoryear{{O'Brien}, {Freeman}  \& {van der
  Kruit}}{{O'Brien} et~al.}{2010}]{2010A&A...515A..63O}
{O'Brien} J.~C.,  {Freeman} K.~C.,   {van der Kruit} P.~C.,  2010, \mn@doi
  [\aap] {10.1051/0004-6361/200912568}, \href
  {https://ui.adsabs.harvard.edu/abs/2010A&A...515A..63O} {515, A63}

\bibitem[\protect\citeauthoryear{{Olling} \& {van Gorkom}}{{Olling} \& {van
  Gorkom}}{1995}]{1995AIPC..336..121O}
{Olling} R.~P.,  {van Gorkom} J.~H.,  1995, in {Holt} S.~S.,  {Bennett} C.~L.,
  eds,  American Institute of Physics Conference Series Vol. 336, Dark Matter.
  pp 121--124, \mn@doi{10.1063/1.48318}

\bibitem[\protect\citeauthoryear{{Orkney} et~al.,}{{Orkney}
  et~al.}{2021}]{orkney}
{Orkney} M. D.~A.,  et~al., 2021, \mn@doi [\mnras] {10.1093/mnras/stab1066},
  \href {https://ui.adsabs.harvard.edu/abs/2021MNRAS.504.3509O} {504, 3509}

\bibitem[\protect\citeauthoryear{{Padilla} \& {Strauss}}{{Padilla} \&
  {Strauss}}{2008}]{2008MNRAS.388.1321P}
{Padilla} N.~D.,  {Strauss} M.~A.,  2008, \mn@doi [\mnras]
  {10.1111/j.1365-2966.2008.13480.x}, \href
  {http://adsabs.harvard.edu/abs/2008MNRAS.388.1321P} {388, 1321}

\bibitem[\protect\citeauthoryear{{Pato}, {Agertz}, {Bertone}, {Moore}  \&
  {Teyssier}}{{Pato} et~al.}{2010}]{2010PhRvD..82b3531P}
{Pato} M.,  {Agertz} O.,  {Bertone} G.,  {Moore} B.,   {Teyssier} R.,  2010,
  \mn@doi [\prd] {10.1103/PhysRevD.82.023531}, \href
  {http://adsabs.harvard.edu/abs/2010PhRvD..82b3531P} {82, 023531}

\bibitem[\protect\citeauthoryear{{Peebles}}{{Peebles}}{1982}]{1982ApJ...263L...1P}
{Peebles} P.~J.~E.,  1982, \mn@doi [\apjl] {10.1086/183911}, \href
  {https://ui.adsabs.harvard.edu/abs/1982ApJ...263L...1P} {263, L1}

\bibitem[\protect\citeauthoryear{{Peter}, {Rocha}, {Bullock}  \&
  {Kaplinghat}}{{Peter} et~al.}{2013}]{2013MNRAS.430..105P}
{Peter} A. H.~G.,  {Rocha} M.,  {Bullock} J.~S.,   {Kaplinghat} M.,  2013,
  \mn@doi [\mnras] {10.1093/mnras/sts535}, \href
  {https://ui.adsabs.harvard.edu/abs/2013MNRAS.430..105P} {430, 105}

\bibitem[\protect\citeauthoryear{{Peters}, {van der Kruit}, {Allen}  \&
  {Freeman}}{{Peters} et~al.}{2017}]{2017MNRAS.464...65P}
{Peters} S.~P.~C.,  {van der Kruit} P.~C.,  {Allen} R.~J.,   {Freeman} K.~C.,
  2017, \mn@doi [\mnras] {10.1093/mnras/stw2101}, \href
  {https://ui.adsabs.harvard.edu/abs/2017MNRAS.464...65P} {464, 65}

\bibitem[\protect\citeauthoryear{{Planck Collaboration} et~al.,}{{Planck
  Collaboration} et~al.}{2014}]{2014A&A...571A..16P}
{Planck Collaboration} et~al., 2014, \mn@doi [\aap]
  {10.1051/0004-6361/201321591}, \href
  {https://ui.adsabs.harvard.edu/abs/2014A&A...571A..16P} {571, A16}

\bibitem[\protect\citeauthoryear{{Pontzen} \& {Governato}}{{Pontzen} \&
  {Governato}}{2013}]{2013MNRAS.430..121P}
{Pontzen} A.,  {Governato} F.,  2013, \mn@doi [\mnras] {10.1093/mnras/sts529},
  \href {https://ui.adsabs.harvard.edu/abs/2013MNRAS.430..121P} {430, 121}

\bibitem[\protect\citeauthoryear{{Pontzen} \& {Tremmel}}{{Pontzen} \&
  {Tremmel}}{2018}]{tangos}
{Pontzen} A.,  {Tremmel} M.,  2018, \mn@doi [\apjs] {10.3847/1538-4365/aac832},
  \href {https://ui.adsabs.harvard.edu/abs/2018ApJS..237...23P} {237, 23}

\bibitem[\protect\citeauthoryear{{Pontzen}, {Ro{\v s}kar}, {Stinson}, {Woods},
  {Reed}, {Coles}  \& {Quinn}}{{Pontzen} et~al.}{2013}]{pynbody}
{Pontzen} A.,  {Ro{\v s}kar} R.,  {Stinson} G.~S.,  {Woods} R.,  {Reed} D.~M.,
  {Coles} J.,   {Quinn} T.~R.,  2013, {pynbody: Astrophysics Simulation
  Analysis for Python}

\bibitem[\protect\citeauthoryear{{Pontzen}, {Read}, {Teyssier}, {Governato},
  {Gualandris}, {Roth}  \& {Devriendt}}{{Pontzen}
  et~al.}{2015}]{2015MNRAS.451.1366P}
{Pontzen} A.,  {Read} J.~I.,  {Teyssier} R.,  {Governato} F.,  {Gualandris} A.,
   {Roth} N.,   {Devriendt} J.,  2015, \mn@doi [\mnras]
  {10.1093/mnras/stv1032}, \href
  {https://ui.adsabs.harvard.edu/abs/2015MNRAS.451.1366P} {451, 1366}

\bibitem[\protect\citeauthoryear{{Pontzen}, {Rey}, {Cadiou}, {Agertz},
  {Teyssier}, {Read}  \& {Orkney}}{{Pontzen} et~al.}{2021}]{pontzen2020}
{Pontzen} A.,  {Rey} M.~P.,  {Cadiou} C.,  {Agertz} O.,  {Teyssier} R.,  {Read}
  J.,   {Orkney} M. D.~A.,  2021, \mn@doi [\mnras] {10.1093/mnras/staa3645},
  \href {https://ui.adsabs.harvard.edu/abs/2021MNRAS.501.1755P} {501, 1755}

\bibitem[\protect\citeauthoryear{{Posti} \& {Helmi}}{{Posti} \&
  {Helmi}}{2019}]{2019A&A...621A..56P}
{Posti} L.,  {Helmi} A.,  2019, \mn@doi [\aap] {10.1051/0004-6361/201833355},
  \href {https://ui.adsabs.harvard.edu/abs/2019A&A...621A..56P} {621, A56}

\bibitem[\protect\citeauthoryear{{Press} \& {Schechter}}{{Press} \&
  {Schechter}}{1974}]{1974ApJ...187..425P}
{Press} W.~H.,  {Schechter} P.,  1974, \mn@doi [\apj] {10.1086/152650}, \href
  {https://ui.adsabs.harvard.edu/abs/1974ApJ...187..425P} {187, 425}

\bibitem[\protect\citeauthoryear{{Read}}{{Read}}{2014}]{2014JPhG...41f3101R}
{Read} J.~I.,  2014, \mn@doi [Journal of Physics G Nuclear Physics]
  {10.1088/0954-3899/41/6/063101}, \href
  {https://ui.adsabs.harvard.edu/abs/2014JPhG...41f3101R} {41, 063101}

\bibitem[\protect\citeauthoryear{{Read} \& {Steger}}{{Read} \&
  {Steger}}{2017}]{2017MNRAS.471.4541R}
{Read} J.~I.,  {Steger} P.,  2017, \mn@doi [\mnras] {10.1093/mnras/stx1798},
  \href {https://ui.adsabs.harvard.edu/abs/2017MNRAS.471.4541R} {471, 4541}

\bibitem[\protect\citeauthoryear{{Read}, {Mayer}, {Brooks}, {Governato}  \&
  {Lake}}{{Read} et~al.}{2009}]{2009MNRAS.397...44R}
{Read} J.~I.,  {Mayer} L.,  {Brooks} A.~M.,  {Governato} F.,   {Lake} G.,
  2009, \mn@doi [\mnras] {10.1111/j.1365-2966.2009.14757.x}, \href
  {https://ui.adsabs.harvard.edu/abs/2009MNRAS.397...44R} {397, 44}

\bibitem[\protect\citeauthoryear{{Read}, {Iorio}, {Agertz}  \&
  {Fraternali}}{{Read} et~al.}{2017}]{2017MNRAS.467.2019R}
{Read} J.~I.,  {Iorio} G.,  {Agertz} O.,   {Fraternali} F.,  2017, \mn@doi
  [\mnras] {10.1093/mnras/stx147}, \href
  {https://ui.adsabs.harvard.edu/abs/2017MNRAS.467.2019R} {467, 2019}

\bibitem[\protect\citeauthoryear{{Read}, {Walker}  \& {Steger}}{{Read}
  et~al.}{2019}]{2019MNRAS.484.1401R}
{Read} J.~I.,  {Walker} M.~G.,   {Steger} P.,  2019, \mn@doi [\mnras]
  {10.1093/mnras/sty3404}, \href
  {https://ui.adsabs.harvard.edu/abs/2019MNRAS.484.1401R} {484, 1401}

\bibitem[\protect\citeauthoryear{{Rey} \& {Pontzen}}{{Rey} \&
  {Pontzen}}{2018}]{2018MNRAS.474...45R}
{Rey} M.~P.,  {Pontzen} A.,  2018, \mn@doi [\mnras] {10.1093/mnras/stx2744},
  \href {https://ui.adsabs.harvard.edu/abs/2018MNRAS.474...45R} {474, 45}

\bibitem[\protect\citeauthoryear{{Rey}, {Pontzen}, {Agertz}, {Orkney}, {Read},
  {Saintonge}  \& {Pedersen}}{{Rey} et~al.}{2019}]{rey2019}
{Rey} M.~P.,  {Pontzen} A.,  {Agertz} O.,  {Orkney} M. D.~A.,  {Read} J.~I.,
  {Saintonge} A.,   {Pedersen} C.,  2019, \mn@doi [\apjl]
  {10.3847/2041-8213/ab53dd}, \href
  {https://ui.adsabs.harvard.edu/abs/2019ApJ...886L...3R} {886, L3}

\bibitem[\protect\citeauthoryear{{Rey}, {Pontzen}, {Agertz}, {Orkney}, {Read}
  \& {Rosdahl}}{{Rey} et~al.}{2020}]{rey2020}
{Rey} M.~P.,  {Pontzen} A.,  {Agertz} O.,  {Orkney} M. D.~A.,  {Read} J.~I.,
  {Rosdahl} J.,  2020, \mn@doi [\mnras] {10.1093/mnras/staa1640}, \href
  {https://ui.adsabs.harvard.edu/abs/2020MNRAS.497.1508R} {497, 1508}

\bibitem[\protect\citeauthoryear{{Rey}, {Katz}, {Cameron}, {Devriendt}  \&
  {Slyz}}{{Rey} et~al.}{2023a}]{2023arXiv230208521R}
{Rey} M.~P.,  {Katz} H.~B.,  {Cameron} A.~J.,  {Devriendt} J.,   {Slyz} A.,
  2023a, arXiv e-prints, \href
  {https://ui.adsabs.harvard.edu/abs/2023arXiv230208521R} {p. arXiv:2302.08521}

\bibitem[\protect\citeauthoryear{{Rey} et~al.,}{{Rey}
  et~al.}{2023b}]{2022arXiv221115689R}
{Rey} M.~P.,  et~al., 2023b, \mn@doi [\mnras] {10.1093/mnras/stad513}, \href
  {https://ui.adsabs.harvard.edu/abs/2023MNRAS.521..995R} {521, 995}

\bibitem[\protect\citeauthoryear{{Roth}, {Pontzen}  \& {Peiris}}{{Roth}
  et~al.}{2016}]{2016MNRAS.455..974R}
{Roth} N.,  {Pontzen} A.,   {Peiris} H.~V.,  2016, \mn@doi [\mnras]
  {10.1093/mnras/stv2375}, \href
  {https://ui.adsabs.harvard.edu/abs/2016MNRAS.455..974R} {455, 974}

\bibitem[\protect\citeauthoryear{{Shao}, {Cautun}, {Frenk}, {Grand},
  {G{\'o}mez}, {Marinacci}  \& {Simpson}}{{Shao}
  et~al.}{2018}]{2018MNRAS.476.1796S}
{Shao} S.,  {Cautun} M.,  {Frenk} C.~S.,  {Grand} R. J.~J.,  {G{\'o}mez} F.~A.,
   {Marinacci} F.,   {Simpson} C.~M.,  2018, \mn@doi [\mnras]
  {10.1093/mnras/sty343}, \href
  {https://ui.adsabs.harvard.edu/abs/2018MNRAS.476.1796S} {476, 1796}

\bibitem[\protect\citeauthoryear{{Simon}}{{Simon}}{2019}]{2019ARA&A..57..375S}
{Simon} J.~D.,  2019, \mn@doi [\araa] {10.1146/annurev-astro-091918-104453},
  \href {https://ui.adsabs.harvard.edu/abs/2019ARA&A..57..375S} {57, 375}

\bibitem[\protect\citeauthoryear{{Sparre} \& {Hansen}}{{Sparre} \&
  {Hansen}}{2012}]{2012JCAP...10..049S}
{Sparre} M.,  {Hansen} S.~H.,  2012, \mn@doi [\jcap]
  {10.1088/1475-7516/2012/10/049}, \href
  {https://ui.adsabs.harvard.edu/abs/2012JCAP...10..049S} {2012, 049}

\bibitem[\protect\citeauthoryear{{Springel}, {White}  \&
  {Hernquist}}{{Springel} et~al.}{2004}]{2004IAUS..220..421S}
{Springel} V.,  {White} S.~D.~M.,   {Hernquist} L.,  2004, in {Ryder} S.,
  {Pisano} D.,  {Walker} M.,   {Freeman} K.,  eds,  IAU Symposium Vol. 220,
  Dark Matter in Galaxies. p.~421

\bibitem[\protect\citeauthoryear{{Stopyra}, {Pontzen}, {Peiris}, {Roth}  \&
  {Rey}}{{Stopyra} et~al.}{2021}]{genetic}
{Stopyra} S.,  {Pontzen} A.,  {Peiris} H.,  {Roth} N.,   {Rey} M.~P.,  2021,
  \mn@doi [\apjs] {10.3847/1538-4365/abcd94}, \href
  {https://ui.adsabs.harvard.edu/abs/2021ApJS..252...28S} {252, 28}

\bibitem[\protect\citeauthoryear{{Teyssier}}{{Teyssier}}{2002}]{2002A&A...385..337T}
{Teyssier} R.,  2002, \mn@doi [\aap] {10.1051/0004-6361:20011817}, \href
  {http://adsabs.harvard.edu/abs/2002A%26A...385..337T} {385, 337}

\bibitem[\protect\citeauthoryear{{Tissera} \& {Dominguez-Tenreiro}}{{Tissera}
  \& {Dominguez-Tenreiro}}{1998}]{1998MNRAS.297..177T}
{Tissera} P.~B.,  {Dominguez-Tenreiro} R.,  1998, \mn@doi [\mnras]
  {10.1046/j.1365-8711.1998.01440.x}, \href
  {https://ui.adsabs.harvard.edu/abs/1998MNRAS.297..177T} {297, 177}

\bibitem[\protect\citeauthoryear{{Tissera}, {White}, {Pedrosa}  \&
  {Scannapieco}}{{Tissera} et~al.}{2010}]{2010MNRAS.406..922T}
{Tissera} P.~B.,  {White} S. D.~M.,  {Pedrosa} S.,   {Scannapieco} C.,  2010,
  \mn@doi [\mnras] {10.1111/j.1365-2966.2010.16777.x}, \href
  {https://ui.adsabs.harvard.edu/abs/2010MNRAS.406..922T} {406, 922}

\bibitem[\protect\citeauthoryear{{Tomassetti} et~al.,}{{Tomassetti}
  et~al.}{2016}]{2016MNRAS.458.4477T}
{Tomassetti} M.,  et~al., 2016, \mn@doi [\mnras] {10.1093/mnras/stw606}, \href
  {http://adsabs.harvard.edu/abs/2016MNRAS.458.4477T} {458, 4477}

\bibitem[\protect\citeauthoryear{{Tormen}}{{Tormen}}{1997}]{1997MNRAS.290..411T}
{Tormen} G.,  1997, \mn@doi [\mnras] {10.1093/mnras/290.3.411}, \href
  {https://ui.adsabs.harvard.edu/abs/1997MNRAS.290..411T} {290, 411}

\bibitem[\protect\citeauthoryear{{Tormen}, {Bouchet}  \& {White}}{{Tormen}
  et~al.}{1997}]{1997MNRAS.286..865T}
{Tormen} G.,  {Bouchet} F.~R.,   {White} S.~D.~M.,  1997, \mn@doi [\mnras]
  {10.1093/mnras/286.4.865}, \href
  {https://ui.adsabs.harvard.edu/abs/1997MNRAS.286..865T} {286, 865}

\bibitem[\protect\citeauthoryear{{Treu}}{{Treu}}{2010}]{2010ARA&A..48...87T}
{Treu} T.,  2010, \mn@doi [\araa] {10.1146/annurev-astro-081309-130924}, \href
  {https://ui.adsabs.harvard.edu/abs/2010ARA&A..48...87T} {48, 87}

\bibitem[\protect\citeauthoryear{{Udry} \& {Martinet}}{{Udry} \&
  {Martinet}}{1994}]{1994A&A...281..314U}
{Udry} S.,  {Martinet} L.,  1994, \aap, \href
  {https://ui.adsabs.harvard.edu/abs/1994A&A...281..314U} {281, 314}

\bibitem[\protect\citeauthoryear{{Vargya}, {Sanderson}, {Sameie},
  {Boylan-Kolchin}, {Hopkins}, {Wetzel}  \& {Graus}}{{Vargya}
  et~al.}{2022}]{2022MNRAS.516.2389V}
{Vargya} D.,  {Sanderson} R.,  {Sameie} O.,  {Boylan-Kolchin} M.,  {Hopkins}
  P.~F.,  {Wetzel} A.,   {Graus} A.,  2022, \mn@doi [\mnras]
  {10.1093/mnras/stac2069}, \href
  {https://ui.adsabs.harvard.edu/abs/2022MNRAS.516.2389V} {516, 2389}

\bibitem[\protect\citeauthoryear{{Vasiliev}, {Belokurov}  \&
  {Evans}}{{Vasiliev} et~al.}{2022}]{2022ApJ...926..203V}
{Vasiliev} E.,  {Belokurov} V.,   {Evans} N.~W.,  2022, \mn@doi [\apj]
  {10.3847/1538-4357/ac4fbc}, \href
  {https://ui.adsabs.harvard.edu/abs/2022ApJ...926..203V} {926, 203}

\bibitem[\protect\citeauthoryear{{Vera-Ciro} \& {Helmi}}{{Vera-Ciro} \&
  {Helmi}}{2013}]{2013ApJ...773L...4V}
{Vera-Ciro} C.,  {Helmi} A.,  2013, \mn@doi [\apjl]
  {10.1088/2041-8205/773/1/L4}, \href
  {https://ui.adsabs.harvard.edu/abs/2013ApJ...773L...4V} {773, L4}

\bibitem[\protect\citeauthoryear{{Vera-Ciro}, {Sales}, {Helmi}, {Frenk},
  {Navarro}, {Springel}, {Vogelsberger}  \& {White}}{{Vera-Ciro}
  et~al.}{2011}]{2011MNRAS.416.1377V}
{Vera-Ciro} C.~A.,  {Sales} L.~V.,  {Helmi} A.,  {Frenk} C.~S.,  {Navarro}
  J.~F.,  {Springel} V.,  {Vogelsberger} M.,   {White} S. D.~M.,  2011, \mn@doi
  [\mnras] {10.1111/j.1365-2966.2011.19134.x}, \href
  {https://ui.adsabs.harvard.edu/abs/2011MNRAS.416.1377V} {416, 1377}

\bibitem[\protect\citeauthoryear{{Vitral} \& {Boldrini}}{{Vitral} \&
  {Boldrini}}{2022}]{2021arXiv211201265V}
{Vitral} E.,  {Boldrini} P.,  2022, \mn@doi [\aap]
  {10.1051/0004-6361/202244530}, \href
  {https://ui.adsabs.harvard.edu/abs/2022A&A...667A.112V} {667, A112}

\bibitem[\protect\citeauthoryear{{Warren}, {Quinn}, {Salmon}  \&
  {Zurek}}{{Warren} et~al.}{1992}]{1992ApJ...399..405W}
{Warren} M.~S.,  {Quinn} P.~J.,  {Salmon} J.~K.,   {Zurek} W.~H.,  1992,
  \mn@doi [\apj] {10.1086/171937}, \href
  {http://adsabs.harvard.edu/abs/1992ApJ...399..405W} {399, 405}

\bibitem[\protect\citeauthoryear{{Weijmans}, {Krajnovi{\'c}}, {van de Ven},
  {Oosterloo}, {Morganti}  \& {de Zeeuw}}{{Weijmans}
  et~al.}{2008}]{2008MNRAS.383.1343W}
{Weijmans} A.-M.,  {Krajnovi{\'c}} D.,  {van de Ven} G.,  {Oosterloo} T.~A.,
  {Morganti} R.,   {de Zeeuw} P.~T.,  2008, \mn@doi [\mnras]
  {10.1111/j.1365-2966.2007.12680.x}, \href
  {https://ui.adsabs.harvard.edu/abs/2008MNRAS.383.1343W} {383, 1343}

\bibitem[\protect\citeauthoryear{{White} \& {Rees}}{{White} \&
  {Rees}}{1978}]{1978MNRAS.183..341W}
{White} S.~D.~M.,  {Rees} M.~J.,  1978, \mn@doi [\mnras]
  {10.1093/mnras/183.3.341}, \href
  {https://ui.adsabs.harvard.edu/abs/1978MNRAS.183..341W} {183, 341}

\bibitem[\protect\citeauthoryear{{Wu} \& {Zhang}}{{Wu} \&
  {Zhang}}{2017}]{2018AAS...23135606W}
{Wu} P.,  {Zhang} S.,  2017, \mn@doi [arXiv e-prints]
  {10.48550/arXiv.1711.09308}, \href
  {https://ui.adsabs.harvard.edu/abs/2017arXiv171109308W} {p. arXiv:1711.09308}

\bibitem[\protect\citeauthoryear{{Zait}, {Hoffman}  \& {Shlosman}}{{Zait}
  et~al.}{2008}]{2008ApJ...682..835Z}
{Zait} A.,  {Hoffman} Y.,   {Shlosman} I.,  2008, \mn@doi [\apj]
  {10.1086/589431}, \href
  {https://ui.adsabs.harvard.edu/abs/2008ApJ...682..835Z} {682, 835}

\bibitem[\protect\citeauthoryear{{Zemp}, {Gnedin}, {Gnedin}  \&
  {Kravtsov}}{{Zemp} et~al.}{2011}]{2011ApJS..197...30Z}
{Zemp} M.,  {Gnedin} O.~Y.,  {Gnedin} N.~Y.,   {Kravtsov} A.~V.,  2011, \mn@doi
  [\apjs] {10.1088/0067-0049/197/2/30}, \href
  {http://adsabs.harvard.edu/abs/2011ApJS..197...30Z} {197, 30}

\bibitem[\protect\citeauthoryear{{Zhang} et~al.,}{{Zhang}
  et~al.}{2019}]{2019MNRAS.484.5170Z}
{Zhang} H.,  et~al., 2019, \mn@doi [\mnras] {10.1093/mnras/stz339}, \href
  {https://ui.adsabs.harvard.edu/abs/2019MNRAS.484.5170Z} {484, 5170}

\bibitem[\protect\citeauthoryear{{Zhu}, {Hernquist}, {Marinacci}, {Springel}
  \& {Li}}{{Zhu} et~al.}{2017}]{2017MNRAS.466.3876Z}
{Zhu} Q.,  {Hernquist} L.,  {Marinacci} F.,  {Springel} V.,   {Li} Y.,  2017,
  \mn@doi [\mnras] {10.1093/mnras/stw3387}, \href
  {https://ui.adsabs.harvard.edu/abs/2017MNRAS.466.3876Z} {466, 3876}

\bibitem[\protect\citeauthoryear{{Zolotov}, {Willman}, {Brooks}, {Governato},
  {Brook}, {Hogg}, {Quinn}  \& {Stinson}}{{Zolotov}
  et~al.}{2009}]{2009ApJ...702.1058Z}
{Zolotov} A.,  {Willman} B.,  {Brooks} A.~M.,  {Governato} F.,  {Brook} C.~B.,
  {Hogg} D.~W.,  {Quinn} T.,   {Stinson} G.,  2009, \mn@doi [\apj]
  {10.1088/0004-637X/702/2/1058}, \href
  {https://ui.adsabs.harvard.edu/abs/2009ApJ...702.1058Z} {702, 1058}

\bibitem[\protect\citeauthoryear{{Zolotov} et~al.,}{{Zolotov}
  et~al.}{2015}]{2015MNRAS.450.2327Z}
{Zolotov} A.,  et~al., 2015, \mn@doi [\mnras] {10.1093/mnras/stv740}, \href
  {http://adsabs.harvard.edu/abs/2015MNRAS.450.2327Z} {450, 2327}

\bibitem[\protect\citeauthoryear{{Zoutendijk} et~al.,}{{Zoutendijk}
  et~al.}{2021}]{2021arXiv211209374Z}
{Zoutendijk} S.~L.,  et~al., 2021, arXiv e-prints, \href
  {https://ui.adsabs.harvard.edu/abs/2021arXiv211209374Z} {p. arXiv:2112.09374}

\bibitem[\protect\citeauthoryear{{van Uitert}, {Hoekstra}, {Schrabback},
  {Gilbank}, {Gladders}  \& {Yee}}{{van Uitert}
  et~al.}{2012}]{2012A&A...545A..71V}
{van Uitert} E.,  {Hoekstra} H.,  {Schrabback} T.,  {Gilbank} D.~G.,
  {Gladders} M.~D.,   {Yee} H.~K.~C.,  2012, \mn@doi [\aap]
  {10.1051/0004-6361/201219295}, \href
  {https://ui.adsabs.harvard.edu/abs/2012A&A...545A..71V} {545, A71}

\makeatother
\end{thebibliography}



\appendix

\section{Shape algorithm} \label{appendix:a}

\begin{figure}
\centering
  \setlength\tabcolsep{2pt}%
    \includegraphics[keepaspectratio, trim={0cm 0cm 0cm 0cm}, width=\columnwidth]{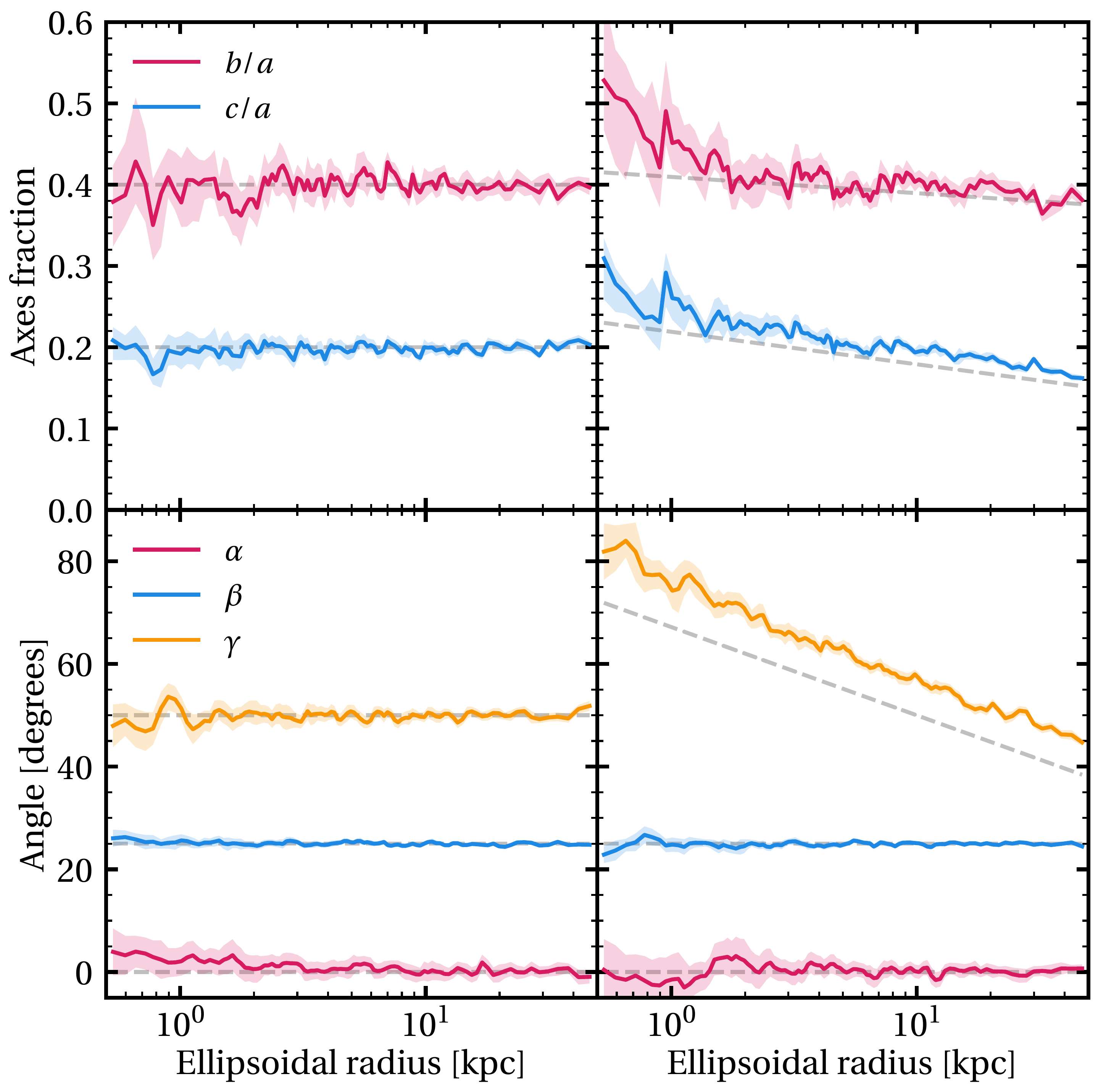}\\
\caption{Shape fits using the algorithm described in \S\ref{method} to mock Hernquist profiles of a known shape. The profile is sampled by $100,000$ equal mass particles. Grey dashed lines indicate the true shape of the profiles. The upper panels show the axis ratio, and the lower panels the corresponding Euler angles as derived from the rotation matrices. These shape parameters are plotted over an arbitrary range of radii in 100 bins, where each bin (initially) contains an equal number of particles. The left panels show the fit to a Hernquist profile with constant axis ratios and rotation over all radii, whereas we vary these parameters in the right panels. The shaded regions correspond to the $1\sigma$ scatter from a bootstrap method.}
\label{fig:shapetest}
\end{figure}

We apply our shape algorithm to four example Hernquist \citep{1990ApJ...356..359H} particle distributions with known axial ratios and orientations in Figure \ref{fig:shapetest}. The fits show good adherence to the underlying particle distribution when the shape parameters are held constant with radius. However, there is a small but consistent systematic bias when the shape parameters vary as a function of radius. This bias is independent of the bin width. \citet{2011ApJS..197...30Z} report a similar bias, which appears to be an innate weakness of the algorithm, and we defer further improvement to future work. \par

The algorithm is included in the {\sc pynbody} \citep{pynbody} halo analysis functions. \par

\section{Resolution study} \label{appendix:c}

\begin{figure*}
\centering
  \setlength\tabcolsep{2pt}%
    \includegraphics[keepaspectratio, trim={0.3cm 0cm 0.2cm 0cm}, width=\linewidth]{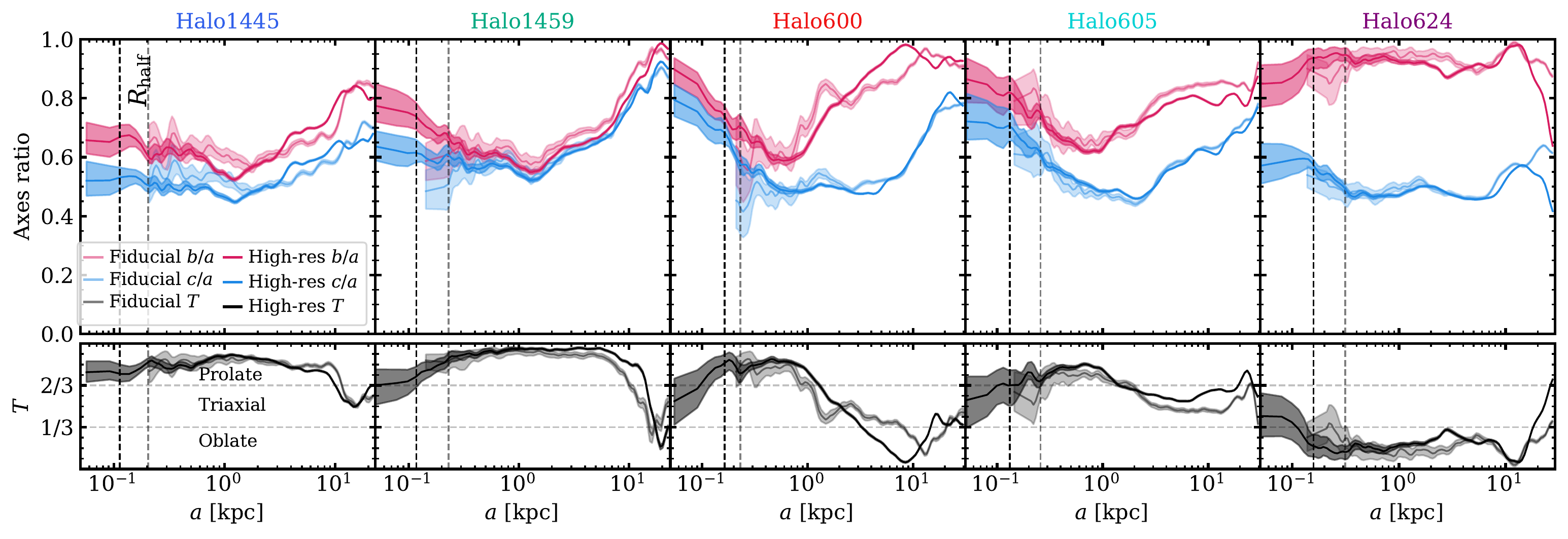}\\
\caption{The same as Figure \ref{fig:shape}, but comparing fiducial resolution (lighter lines) with high-resolution (darker lines) simulation versions. The qualitative form of the halo shape is insensitive to the simulation resolution, except towards the resolution limit of the ``fiducial'' resolution versions.}
\label{fig:shape_res}
\end{figure*}

\begin{figure*}
\centering
  \setlength\tabcolsep{2pt}%
    \includegraphics[keepaspectratio, trim={0.3cm 0cm 0.2cm 0cm}, width=\linewidth]{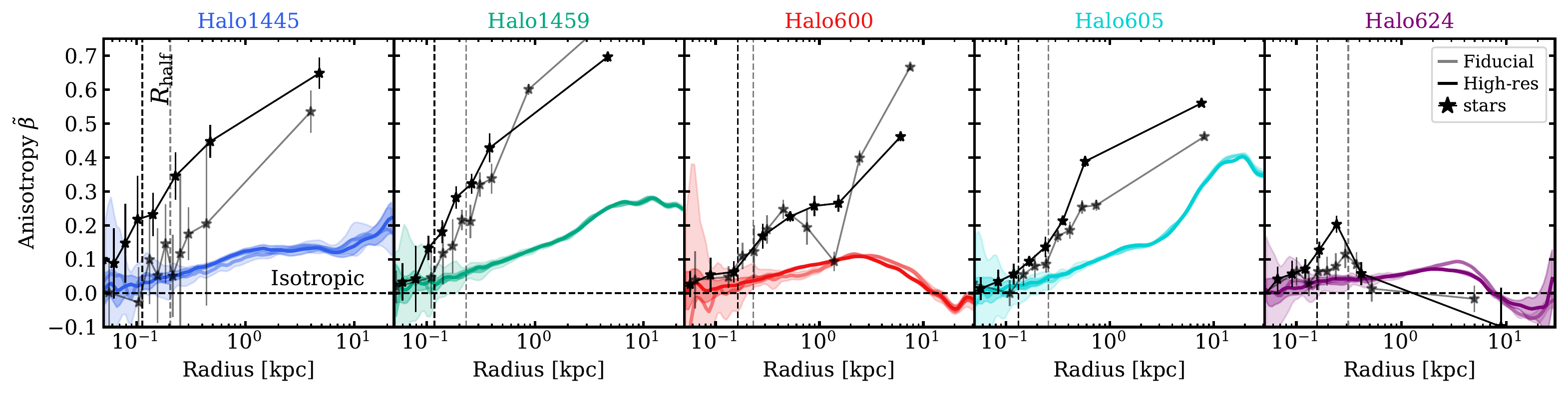}\\
\caption{The same as Figure \ref{fig:anisotropy}, but comparing fiducial resolution (lighter lines) with high-resolution (darker lines) simulation versions. The velocity anisotropy of the \ac{DM} is not sensitive to resolution, but the stellar velocity anisotropy is.}
\label{fig:anisotropy_res}
\end{figure*}

In this section, we perform a comparison of the axial ratio and velocity anisotropy profiles between two different simulation resolutions, hereafter `fiducial' and `high-res'. The fiducial resolution versions of each halo are simulated with \ac{DM} particles up to a maximum resolution of $960\,\rm{M}_{\odot}$ in the zoom region, whereas the high-res versions are simulated with \ac{DM} particles up to a maximum resolution of $120\,\rm{M}_{\odot}$. In both cases, we compare simulations with baryonic physics where the gas elements are allowed to refine to a minimum comoving length of 3\,pc. See \citet{2023arXiv230208521R} for an investigation into how computational gas grid resolutions can affect the propagation of gas flows. \par

As shown in Figure \ref{fig:shape_res}, the shape profiles are broadly comparable between each resolution, with small discrepancies arising at various radii. The fiducial resolution versions do not contain sufficient particles to estimate the halo shape within $\sim0.1\,$kpc, and the agreement of the shape fit is particularly poor for radii approaching this inner limit. However, they remain roughly consistent if the $1\sigma$ uncertainties are accounted for. \par

There are some further disagreements between each resolution at larger radii (i.e. $\approx10$\,kpc), with the differences exceeding the mutual $1\sigma$ uncertainties. This suggests that differences in the axial ratios at this scale (within about $\Delta(x/a)=0.1$, with $x=b,c$) should not be relied upon too strongly when interpreting results. \par

The velocity anisotropy profiles in Figure \ref{fig:anisotropy_res} show exeedingly close alignment between the \ac{DM} profiles in both resolution versions, although the temporal uncertainty is far greater at $r<0.5\,$kpc in the fiducial resolution versions. There are similar trends for the stellar velocity anisotropy too, although the profiles are systematically lower in the fiducial resolution versions. Following the reasoning in \S\ref{stars_DM}, this is likely because merging haloes are more rapidly disrupted in the fiducial resolution simulations, and so the stellar profiles are less contaminated by radially-biased ex-situ stars at increasingly low radii. The co-rotating stellar features reported in Halo600 and Halo624 are also present in the fiducial resolution versions. \par

\section{Merger trees} \label{appendix:b}

\begin{figure*}
\centering
  \setlength\tabcolsep{2pt}%
    \includegraphics[keepaspectratio, trim={0cm 0cm 0cm 0cm}, width=0.8\linewidth]{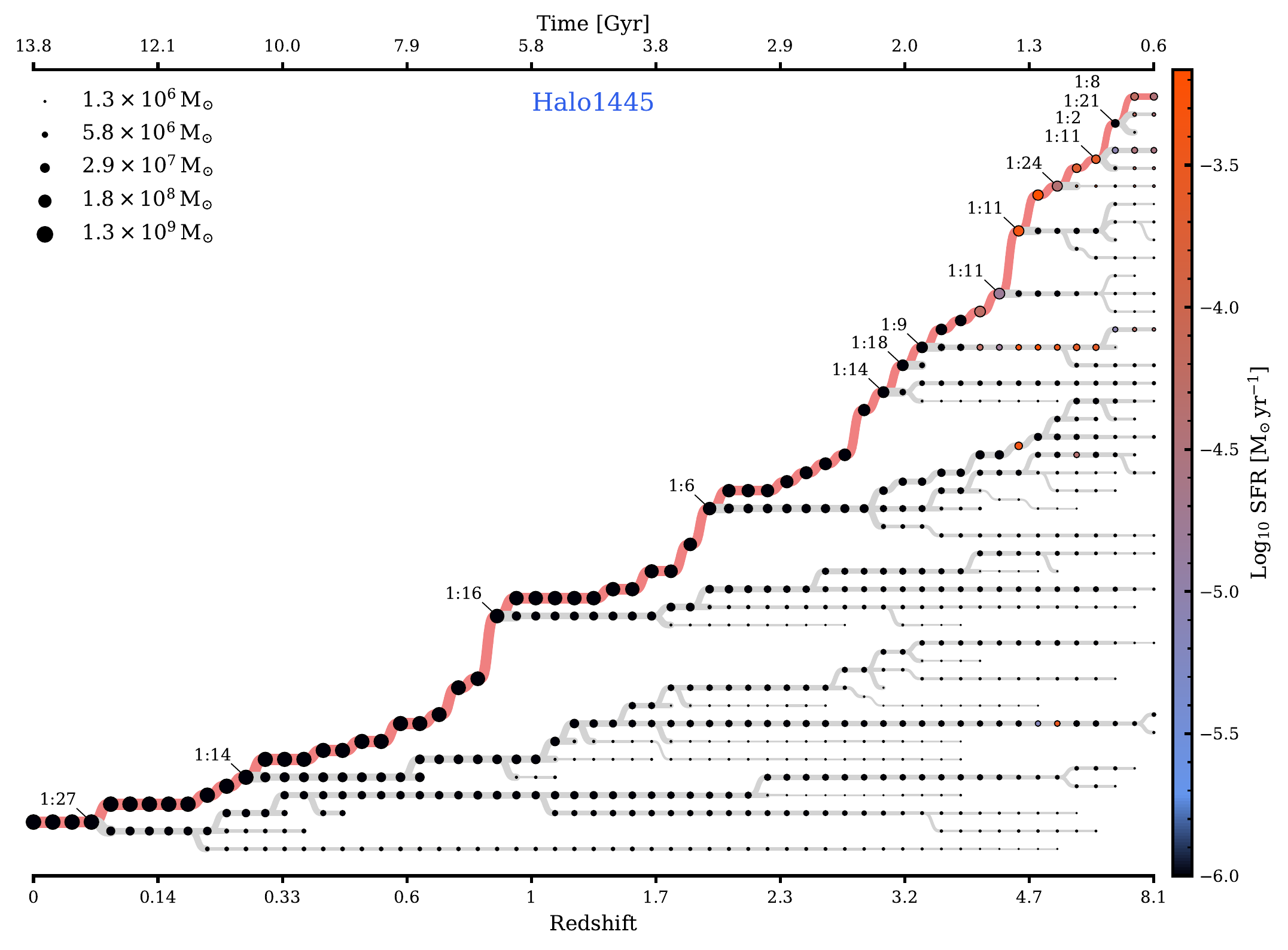}\\
\caption{Merger tree visualisation for the baryonic version of Halo1445. The main progenitor undergoes several significant mergers across cosmic time.}
\label{fig:tree_1445}
\end{figure*}

\begin{figure*}
\centering
  \setlength\tabcolsep{2pt}%
    \includegraphics[keepaspectratio, trim={0cm 0cm 0cm 0cm}, width=0.8\linewidth]{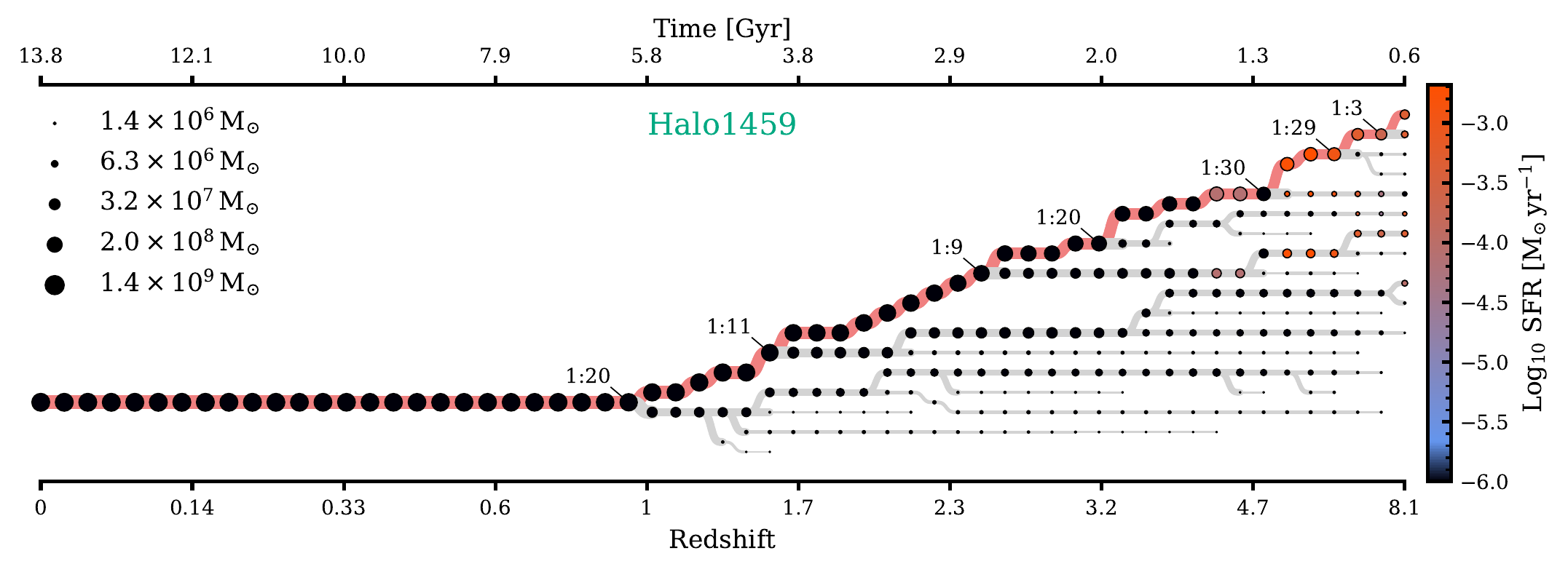}\\
\caption{Merger tree visualisation for the baryonic version of Halo1459. The main progenitor undergoes only a few moderate mergers, all occurring before $z=1$.}
\label{fig:tree_1459}
\end{figure*}

\begin{figure*}
\centering
  \setlength\tabcolsep{2pt}%
    \includegraphics[keepaspectratio, trim={0cm 0cm 0cm 0cm}, width=0.8\linewidth]{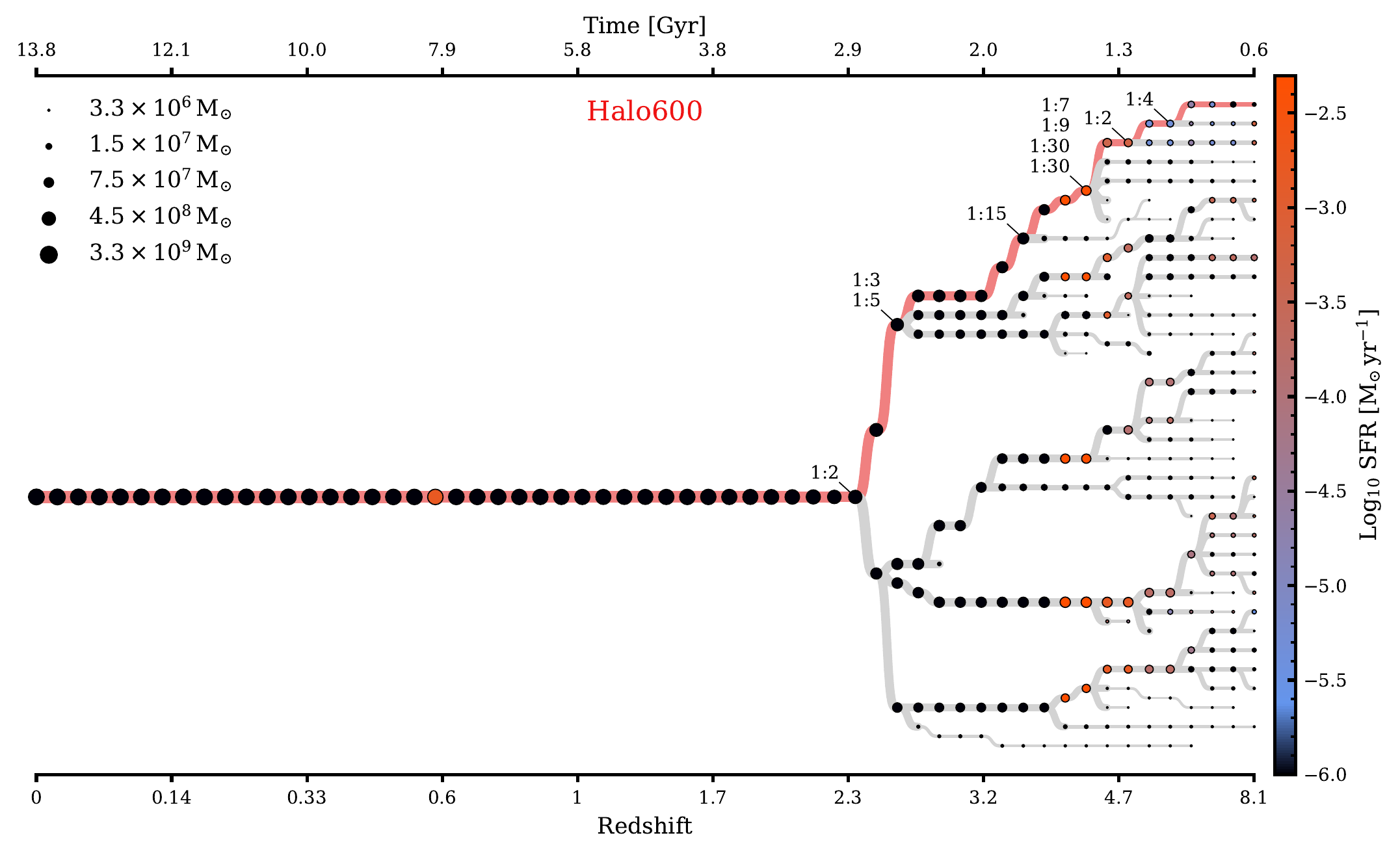}\\
\caption{Merger tree visualisation for the baryonic version of Halo600. This halo assembles rapidly at around 2-3\,Gyr from a convergence of many similar-mass haloes. Each component halo is unable to form a large number of stars due to its relatively low gravitational potential, which is why Halo600 has a lower stellar mass compared to the galaxies at similar final halo mass (Halo605 and Halo624). The rapid assembly at 3\,Gyr causes a sudden change in the \ac{DM} shape (see Figure \ref{fig:shape_with_time}), and kickstarts a process of gas accretion that culminates in a violent starburst at 8\,Gyr.}
\label{fig:tree_600}
\end{figure*}

\begin{figure*}
\centering
  \setlength\tabcolsep{2pt}%
    \includegraphics[keepaspectratio, trim={0cm 0cm -0.26cm 0cm}, width=0.8\linewidth]{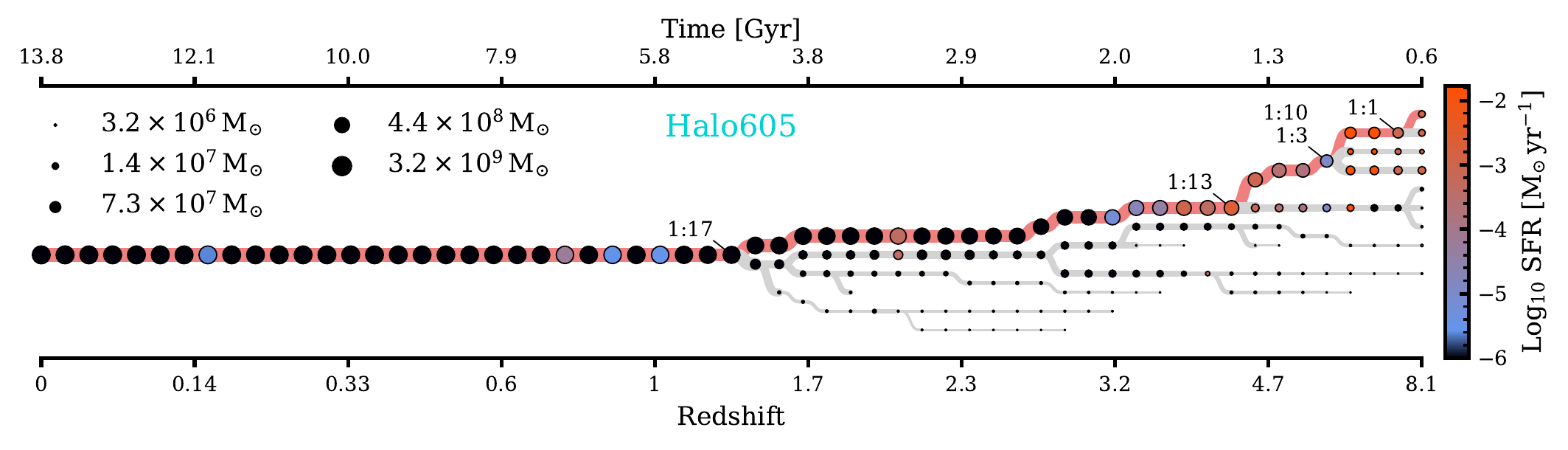}\\
\caption{Merger tree visualisation for the baryonic version of Halo605. The main progenitor undergoes only a few minor mergers, all occurring before $z=1$.}
\label{fig:tree_605}
\end{figure*}

\begin{figure*}
\centering
  \setlength\tabcolsep{2pt}%
    \includegraphics[keepaspectratio, trim={0cm 0cm 0cm 0cm}, width=0.8\linewidth]{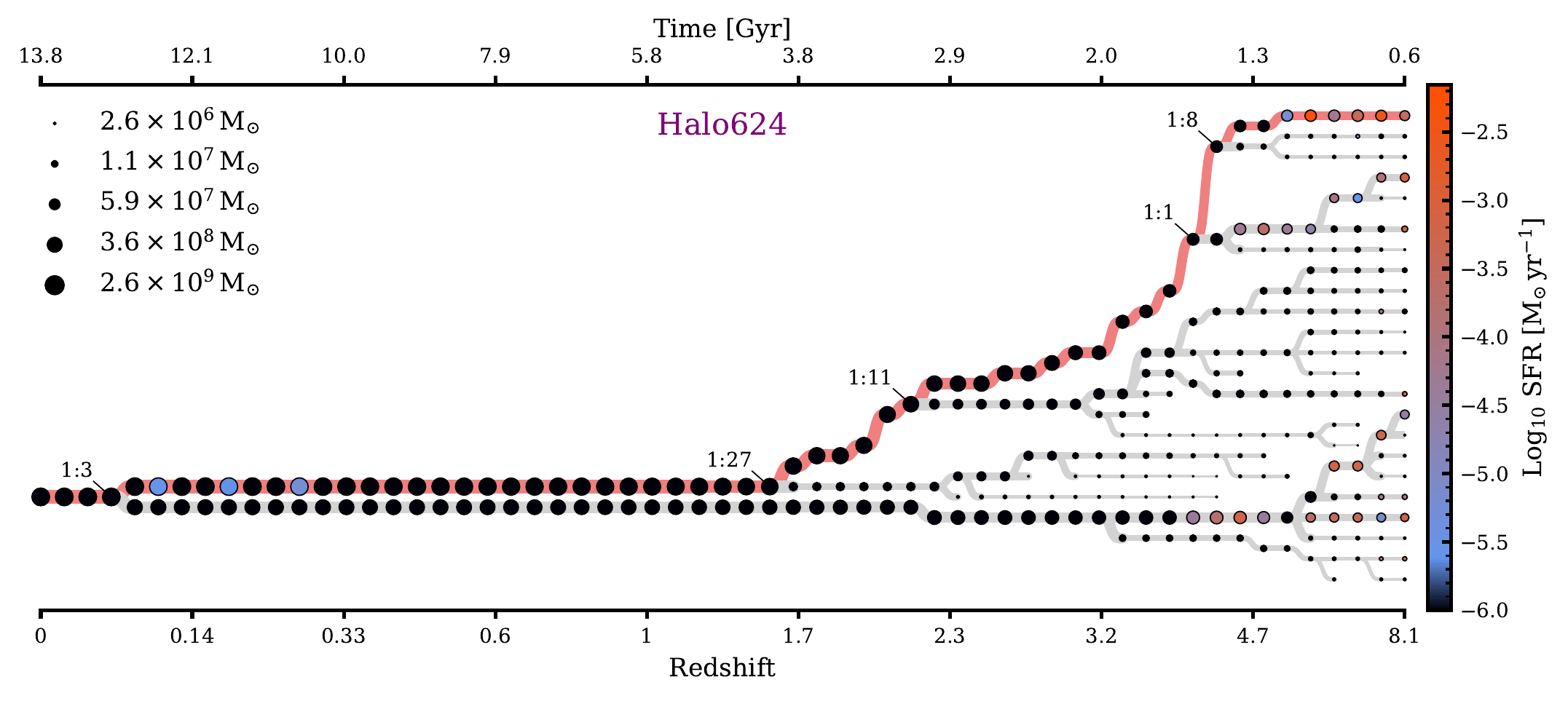}\\
\caption{Merger tree visualisation for the baryonic version of Halo624. Note the near equal-mass merger that occurs after 1.2\,Gyr, which is correlated with a sudden change in the halo shape (see Figure \ref{fig:tree_624}).}
\label{fig:tree_624}
\end{figure*}

\begin{figure*}
\centering
  \setlength\tabcolsep{2pt}%
    \includegraphics[keepaspectratio, trim={0cm 0cm 0cm 0cm}, width=0.8\linewidth]{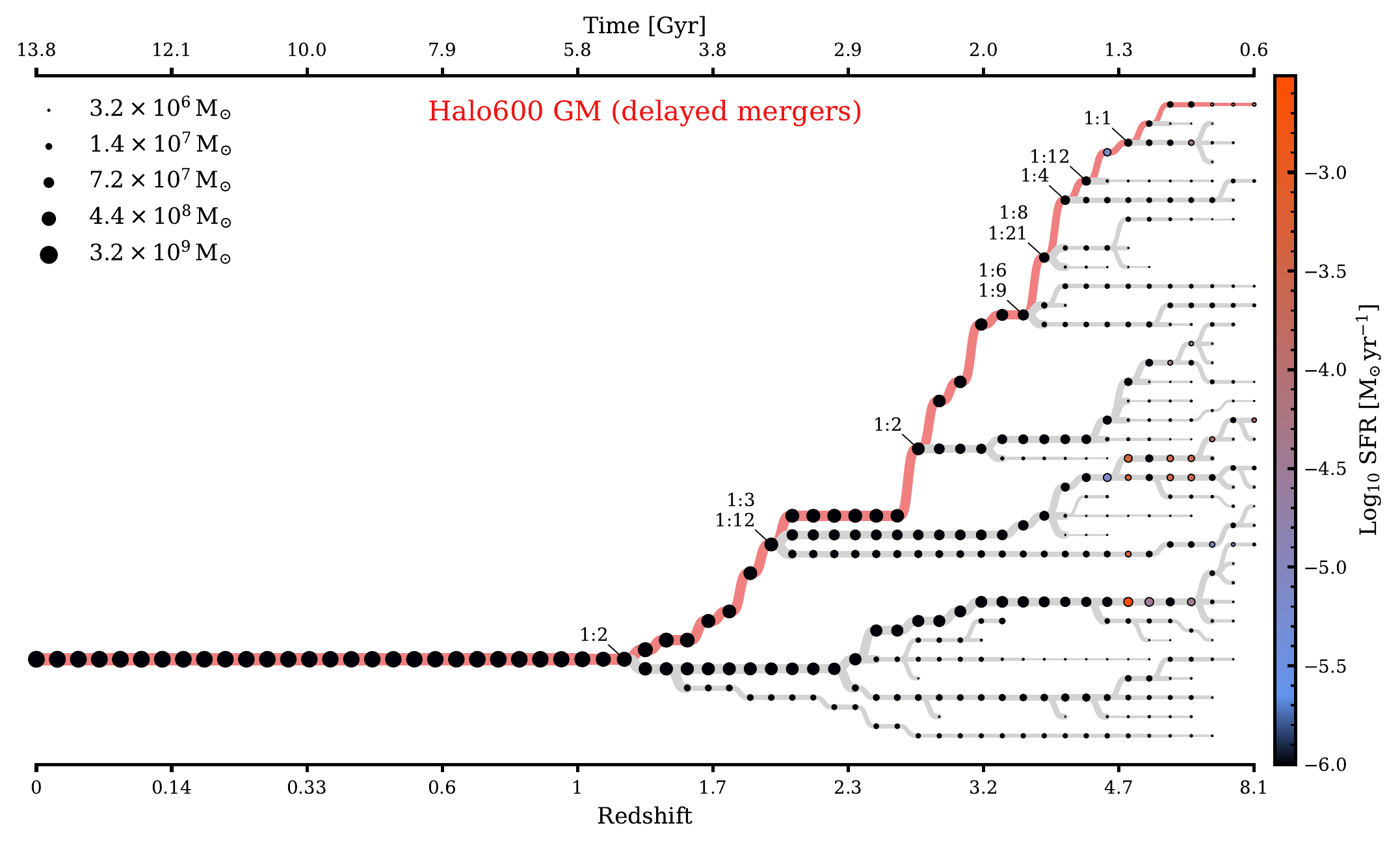}\\
\caption{Merger tree visualisation for the baryonic version of Halo600 GM, a variant of Halo600 that has been `genetically modified' to alter the assembly history. In this case, the merger events occurring at $\sim3\,$Gyr are delayed and spread over a longer time period. The final halo mass is unaltered.}
\label{fig:tree_600_GM}
\end{figure*}

\begin{figure*}
\centering
  \setlength\tabcolsep{2pt}%
    \includegraphics[keepaspectratio, trim={0cm 0cm 0cm 0cm}, width=0.8\linewidth]{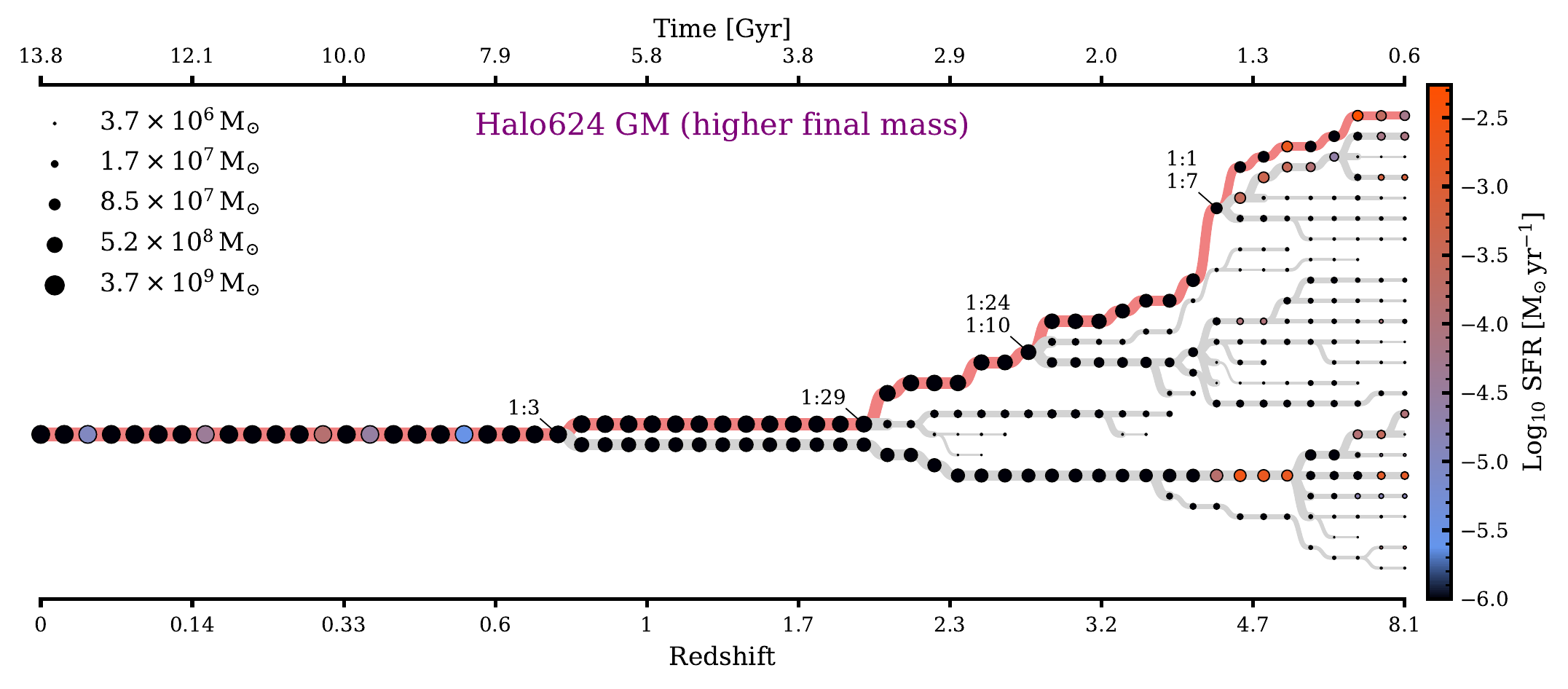}\\
\caption{Merger tree visualisation for the baryonic version of Halo624 GM, a variant of Halo624 that has been `genetically modified' to alter the assembly history. In this case, the final halo mass has been increased by $\sim10^9\,\rm{M}_{\odot}$ and several major mergers now occur at an earlier time.}
\label{fig:tree_624_GM}
\end{figure*}

The merger history of each \ac{EDGE} galaxy is tracked using the tools in {\sc tangos} \citep{tangos}. Here we present visualisations of these merger trees, built with the assistance of {\sc GraphViz} and the {\sc pydot} package for {\sc Python}. We consider haloes that are $1:30$ mass ratio mergers or greater, and no earlier than $z=9$. Each node represents a halo identified with HOP halo finder \citep{hop}, with the connecting lines indicating the halo descendants (left) and progenitors (right). The main progenitor line, defined as the most massive descendant at each step, is highlighted in pink. Annotations on the main progenitor line represent the merger mass ratios. The colour of each node shows the instantaneous star formation rate within each halo, averaged over a time interval of 100\,Myr. \par

We show the merger history of each simulation represented in Table \ref{tab:edge_sims}, along with merger history of the modified fiducial simulations in \S\ref{GM}. We do not show merger histories for the simulations at our lower `fiducial' resolution, or of the weak feedback models in \S\ref{fblim}, because this change in resolution and physics model have negligible impact on the assembly history. \par

We have omitted roughly one in every two snapshots after a redshift $z=1.5$. This is because the merger histories become far more quiescent at later times, and also because each simulation would otherwise have a slightly different number of total outputs. By enforcing a strict selection of outputs, we ensure that each tree is directly comparable and that emphasis is drawn to the merger-rich early epoch. \par 

Here, an accreting halo is ``merged'' once it passes within the virial radius of its host, but will still persist as an independent subhalo for some time after this. There is an unspecified lag time between the infall and the eventual dissolution of the merger, which is not reflected in these figures. \par


\bsp	
\label{lastpage}
\end{document}